\documentclass[preprint2]{aastex}

\usepackage{amssymb}
\usepackage{natbib}
\usepackage{rotating}

\newcommand{\myarcsec}{\hbox{$.\!\!^{\prime\prime}$}}

\newcommand{\AAb}{\AA$\;$}

\newcommand{\kms}{\rm ~km~s^{-1}}

\newcommand{\ergs}{\rm ~erg~s^{-1}}
\newcommand{\ergsm}{\rm ~erg~s^{-1} cm^{-2}}

\DeclareRobustCommand{\ion}[2]{\textup{#1\,\textsc{\lowercase{#2}}}}
\newcommand*\element[1][]{%
  \def\aa@element@tr{#1}%
  \aa@element
}

\shorttitle{Ultra-luminous NLRs - Quasar light echos?}
\shortauthors{Schirmer et al.}

\begin{document}
\bibliographystyle{aa}


\renewcommand{\thefootnote}{\fnsymbol{footnote}}

\title{A sample of Seyfert-2 galaxies with ultra-luminous galaxy-wide NLRs -- Quasar light echos?
\footnote{
  {\bf Based on observations made with ESO Telescopes} at the La Silla and Paranal 
  Observatories, Chile.
  {\bf Based on observations obtained 
  with MegaPrime/MegaCam}, a joint project of CFHT and CEA/DAPNIA, at 
  the Canada-France-Hawaii Telescope (CFHT) which is operated by the 
  National Research Council (NRC) of Canada, the Institut National des 
  Sciences de l'Univers of the Centre National de la Recherche 
  Scientifique (CNRS) of France, and the University of Hawaii.
  {\bf Based on observations obtained at the Gemini Observatory}, which is operated
  by the Association of Universities for 
  Research in Astronomy, Inc., under a cooperative agreement with the NSF 
  on behalf of the Gemini partnership: the National Science Foundation (United
  States), the Science and Technology Facilities Council (United Kingdom), the
  National Research Council (Canada), CONICYT (Chile), the Australian Research Council
  (Australia), Minist\'{e}rio da Ci\^{e}ncia, Tecnologia e Inova\c{c}\~{a}o (Brazil) 
  and Ministerio de Ciencia, Tecnolog\'{i}a e Innovaci\'{o}n Productiva (Argentina).}
}



\author{M. Schirmer\altaffilmark{1,2}, R. Diaz\altaffilmark{1,3}, K. Holhjem\altaffilmark{4}, 
N. A. Levenson\altaffilmark{1}, C. Winge\altaffilmark{1}}
\affil{\altaffilmark{1}Gemini Observatory, Casilla 603, La Serena, Chile}
\affil{\altaffilmark{2}Argelander-Institut f\"ur Astronomie, Universit\"at Bonn, 
  Auf dem H\"ugel 71, 53121 Bonn, Germany}
\affil{\altaffilmark{3}ICATE, CONICET, Argentina}
\affil{\altaffilmark{4}SOAR Telescope, Casilla 603, La Serena, Chile}





\begin{abstract}
We report the discovery of Seyfert-2 galaxies in SDSS-DR8 with galaxy-wide, 
ultra-luminous narrow-line regions (NLRs) at redshifts $z=0.2-0.6$. With a space density of 
4.4 Gpc$^{-3}$ at $z\sim0.3$, these ``Green Beans'' (GBs) are amongst the rarest objects in the 
Universe. We are witnessing an exceptional and/or short-lived phenomenon in the life cycle of 
AGN. The main focus of this paper is on a detailed analysis of the GB prototype 
galaxy J2240-0927 ($z=0.326$). Its NLR extends over $26\times44$ kpc and is surrounded by an 
extended narrow-line region (ENLR). With a total [\ion{O}{III}]$\lambda5008$ luminosity of 
$(5.7\pm0.9)\times10^{43}$ erg s$^{-1}$, this is one of the most luminous NLR known around any 
type-2 galaxy. Using VLT/XSHOOTER we show that the NLR is powered by an AGN, and we derive 
resolved extinction, density and ionization maps. Gas kinematics is disturbed on a global scale,
and high velocity outflows are absent or faint. This NLR is unlike any other NLR or extended emission 
line region (EELR) known. Spectroscopy with Gemini/GMOS reveals extended, high luminosity 
[\ion{O}{III}] emission also in other GBs. WISE 24$\mu$m luminosities are $5-50$ times lower than predicted 
by the [\ion{O}{III}] fluxes, suggesting that the NLRs reflect earlier, very active quasar states that have 
strongly subsided in less than a galaxies' light crossing time. These light echos, or ionization echos,
are about $100$ times more luminous than any other such echo known to date. X-ray data are needed for 
photo-ionization modeling and to verify the light echos.

\end{abstract}


\section{Introduction}
NLRs are a common sight in active galaxies and explained by the unified AGN model
\citep[see e.g.][]{ant93,bmr12}. UV/X-ray emission from the central black hole engine photo-ionizes the ISM over 
large distances, but may be shielded by neutral gas or dust absorbers. The prototype NLR consists of 
two ionization cones (e.g.\ in NGC 4151 and NGC 5252), which may fragment into several individual clouds 
\citep{etk93,tmw96}. In Seyfert-2 galaxies NLR sizes are a few hundred pc, and can extend up to 5 kpc 
\citep{bjk06}. Emission line FWHMs are on the order of several hundred km s$^{-1}$ and are often richly 
structured. 

Extended narrow-line regions (ENLRs) were first described by \cite{upa87} and have been found around many 
AGN since \citep{dur89}. They extend over 20 kpc or more and have much lower luminosities than NLRs. 
Dynamical FWHMs are low ($\lesssim 50$ km s$^{-1}$), whereas high excitation 
levels reveal AGN as ionizing sources. ENLRs usually follow the galactic rotation.

Extended emission line regions \citep[EELRs,][]{fos89} are found mostly around radio-loud QSOs
and have similar sizes as ENLRs, but very different characteristics. \cite{fus09} find compact clouds 
with low line widths ($<100$ km s$^{-1}$) but moving with high velocities ($\sim500$ km s$^{-1}$). 
Dynamically chaotic structures are common, and morphological links to the host galaxies 
absent. EELRs are likely the result of a merger, kick-starting the QSO engine that ionizes the 
gas and blasts it into the outer surroundings.

\cite{lsk09} find a cloud of ionized gas 20 kpc outside IC 2497. Lacking an apparent AGN in the host galaxy, 
it is interpreted as a light echo from a very active earlier AGN phase. Indeed radio observations by \cite{rgj10} 
reveal hidden AGN features. \cite{kcb12} present 154 galaxies with similar detached clouds at 
redshifts $z\lesssim0.1$, finding evidence for short periods of high AGN activity. 
During these periods the luminosity can change by up to 4 orders of magnitude, as has been shown by 
\cite{sev10} and \cite{kls12} for IC 2497. However, time-scales inferred by \cite{kcb12} are too short to be 
explained by current accretion models. Such light echos therefore provide new insights into the onset and 
shutdown processes of QSO activity on scales of a galaxy's light crossing time ($\sim10^4-10^5$ years).

Galaxy formation models predict AGN-driven large outflows needed to explain properties of the 
interstellar medium and massive galaxies \citep[e.g.][]{sdh05}. NLRs on galaxy scales have 
been found in powerful radio galaxies \citep[HzRGs,][]{nld08}, and ultra-luminous infrared galaxies 
\citep[ULIRGs,][]{has12} at high redshifts ($z=2-3$). Line widths of $700-1400$ km s$^{-1}$ and 
large offsets with respect to the systemic redshift emphasize the outflow character in the ULIRGs with highest 
[\ion{O}{III}] luminosity.

\cite{css09} present a new class of emission-line galaxies, dubbed ``Green Peas'' (GPs), referring to their 
compact size and color in $gri$ images. In a sample of 112 GPs they find 80 star-forming galaxies, 9 Seyfert-1, 
10 Seyfert-2 and 13 transition objects (showing both AGN and star formation features). GPs with AGN-like emission 
characteristics have so far not been studied beyond the initial identification and analysis in \cite{css09},
and differences to NLRs in other Seyfert galaxies have not been found.

It is now established that supermassive black holes (SMBHs) reside in the centers of massive 
galaxies, which must have hosted an AGN in their past. Consequently, mergers of galaxies must 
also lead to the coalescence of SMBHs. Indeed numerous wide-separation binary AGN are known, however systems 
with kpc separations \citep{slg11,cgs12} or less \citep{rtz06,fwe11} are rare and have to be verified 
carefully. Simulations of the pre-coalescence state of SMBH mergers \citep[such as][]{hoq10,kbb12,vvm12}
do not predict extraordinary NLR properties.

\begin{figure*}[t]
  \includegraphics[width=1.0\hsize]{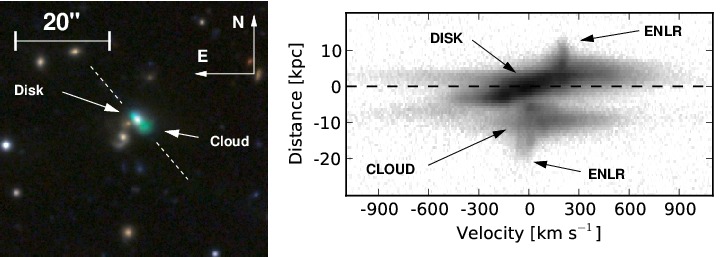}
  \caption{\label{image}{Left: CFHT/Megaprime $gri$ color image of J2240. The photometric redshift of the
      fainter reddish galaxy to the left of the nucleus categorizes it as a background source, whereas the 
      galaxy below with the prominent spiral arm has the same spectroscopic redshift as J2240. The 
      dashed line shows the orientation of the VLT/XSHOOTER slit, running (from the upper left 
      to the lower right) through the Disk and the nucleus, and then through the Cloud.
      Right: Complex appearance of the [\ion{O}{III}]$\lambda5008$ line in the VLT/XSHOOTER spectrum 
      (log-scaled, continuum subtracted). Some of the features discussed in this paper are marked. 
      The dashed line indicates the spatial position of the nucleus.}}
\end{figure*}

In CFHT/Megaprime data we serendipitously discovered J224024.1-092748 (hereafter: J2240), a peculiar 
galaxy at $z=0.326$ with GP colors. Due to its large angular extent of $4^{\prime\prime}\times7^{\prime\prime}$
J2240 was not included in the GP sample of \cite{css09}. In this paper we show that it is fundamentally 
different from GPs. We have identified two dozen similar objects in SDSS-DR8 \citep{aaa11} with redshifts 
$z=0.1-0.7$, after removing 95\% spurious detections. We refer to these galaxies with large and ultra-luminous 
NLRs as ``Green Beans'' (GBs). GBs display a previously unknown phenomenon in the life of AGN, and 
we explore different formation mechanisms.

The paper is structured as follows. In Sect.~2 we present observational 
data for J2240, followed by our 2D spectral analysis method in Sect.~3. We discuss the physical 
properties obtained from this analysis in Sect.~4, after which we present our sample 
of similar galaxies extracted from SDSS. We discuss our findings and summarize in Sect.~6. 
Throughout this paper we assume a flat standard cosmology with $\Omega_m=0.27$, $\Omega_\Lambda=0.73$ and 
$H_0=70\,h_{70}\kms\,{\rm Mpc}^{-1}$. The relation between physical and angular scales 
at $z=0.326$ is $1\myarcsec0 = 4.76\,h_{70}^{-1}\,{\rm kpc}$. Error bars represent the 
$1\sigma$ confidence level. All wavelengths stated are vacuum wavelengths.

\section{Observations}

\subsection{\label{imaging}CFHT imaging}
Deep $ugriz$ CFHT/Megaprime data (Table~\ref{obstable_img}) were obtained 
by us through Opticon proposal 2008BO01 at 2008-09-20 in excellent conditions to study a 
supercluster of galaxies at $z=0.45$. Further details about these data can be found in 
\cite{shk11}. J2240 at $z=0.326$ is a serendipitous discovery in these images, extending over 
$7^{\prime\prime}\times4^{\prime\prime}$. With a pixel scale of 0\farcs186 and 0\farcs7 
image seeing the galaxy is well resolved. It has irregular morphology and peculiar colors similar 
to that of a GP (see Fig.~\ref{image}), with the exception of 
very low $u$-band flux ($u-r=4.06$ mag). K-corrections and luminosities were calculated
using {\tt kcorrect} v4.2 \citep{blr07}. Most noteworthy are a bright non-stellar 
nucleus in a disk-like body reminiscent of a spiral galaxy, and a half-detached 
cloud about 12 kpc south-west of the nucleus. The cloud extends over $8\times18$ kpc.

We hereafter refer to these two components simply as the `Disk' and the `Cloud', without 
implying any properties about their physical shape. Spectroscopic observations reveal a much 
more complex picture, as can be seen in the right panel of Fig.~\ref{image}. We refer to the
central $\pm4$ kpc encompassing the nucleus as the `Center'.

Figure~\ref{image} shows a fainter reddish galaxy with a photometric redshift of $z=0.37\pm0.06$, 
projected 2\myarcsec0 east of the nucleus of J2240. A second galaxy with a prominent spiral 
arm is located 4\myarcsec5 to the south-east, and has the same spectroscopic redshift as J2240 
($z=0.326$).

\subsection{\label{spectroscopy} Spectroscopy}
\subsubsection{VLT/FORS2}
The redshift of J2240 and its classification as a Seyfert-2 galaxy were first secured using
a VLT/FORS2 \citep{aff98} spectrum (ESO DDT proposal 286.A-5027). Our data were reduced using 
a custom pipeline. In short, all spectra were overscan corrected, debiased, flat-fielded, rectified and 
sky subtracted. Flux-calibration and correction for telluric absorption were based on the central 
white dwarf of the Helix Nebula (NGC 7293) and the atmospheric transmission model for Paranal 
\citep{pmo11}. A technical summary of the characteristics of our spectroscopic data sets is given in 
Table~\ref{obstable_spec}.

\begin{table}[t]
\caption{Characteristics of the CFHT/Megaprime data and J2240 photometry. The 
limiting magnitudes represent the $5\sigma$ completeness limit for non-stellar sources.
Luminosities are dereddened.}
\label{obstable_img}
\begin{tabular}{lccc|cc}
\noalign{\smallskip}
\hline
\hline
\noalign{\smallskip}
Band & $t_{\rm exp}$ [s] & Seeing & $M_{\rm lim}$ & $M_{\rm AB}$ & $L$\\
\hline
\noalign{\smallskip}
$u$ & 4250 & 0\myarcsec99 & 25.0 & 21.80 & -19.4\\
$g$ & 3000 & 0\myarcsec76 & 25.4 & 19.02 & -21.3\\
$r$ & 5000 & 0\myarcsec66 & 25.4 & 17.74 & -22.2\\
$i$ & 6000 & 0\myarcsec57 & 25.2 & 18.71 & -23.0\\
$z$ & 5000 & 0\myarcsec73 & 23.3 & 17.89 & -23.2\\
\hline
\end{tabular}
\end{table}

\subsubsection{VLT/XSHOOTER}
Using VLT/XSHOOTER \citep{vdd11}, we obtained a spectrum (proposal 287.B-5008) of the full 
$300-2500$nm range, with a slit position angle of 43.5 degrees (same as for FORS2) aligned along 
the major axis of the extended emission in Fig.~\ref{image}. Spectral (spatial) resolutions 
are $5-20$ (1.5) times higher than with 
FORS2. The data were pre-processed using the XSHOOTER pipeline \citep{mgr10}. Flux 
calibration (using the spectrophotometric standard Feige 110), continuum subtraction and 
further data processing were done using custom-made software. Since the slit width for the
flux standard was much wider than the one for the science target, and the standard was 
calibrated with a different flat field, the absolute calibration is only known to within 
a constant factor. We determine the latter from the FORS2 spectrum, over which we have full 
control and consistent calibrators. After correction, FORS2 and XSHOOTER fluxes are 
indistinguishable within their noise in the common wavelength area. The full XSHOOTER spectrum is
shown in Appendix~\ref{apdx1}, Fig.~\ref{emissionlines}.

\subsubsection{Gemini/GMOS}
We used Gemini-South/GMOS (proposal GS-2012A-Q-83) to create a redshift survey of further 
GB candidates (Sect.~\ref{gbsurvey}). Given that our targets are bright and our only aim 
was to detect bright emission lines, this campaign was designed as a backup program
to be executed in bad seeing and cirrus. Most data were taken in the second half 
of the 2012A semester. At the time of writing observations are still ongoing.

\begin{table*}
\caption{Characteristics of the spectroscopic data. GMOS data were taken throughout the 
2012A semester.}
\label{obstable_spec}
\begin{tabular}{lccccccccc}
\hline
\hline
\noalign{\smallskip}
Instrument & Grating & $\lambda_{\rm min}-\lambda_{\rm max}$ & $t_{\rm exp}$ & Seeing & Slit & 
$\lambda/\Delta\lambda$& Spatial & Spectral\\
 & & [nm] & [s] & [$^{\prime\prime}$] & [$^{\prime\prime}$] & & sampling & sampling & Date\\
 & & & & & & & [$^{\prime\prime}$ pix$^{-1}$] & [\AAb pix$^{-1}$] &\\
\hline
\noalign{\smallskip}
FORS2    & 300V & $445-865$   & $2400$ & 1.0 & 1.0 &  300 & 0.25 & 3.34 & 2010-12-19\\
XSHOOTER & UVB  & $300-560$   & $1160$ & 0.55 & 1.0 & 4350 & 0.16 & 0.15 & 2011-06-13\\
         & VIS  & $550-1020$  & $1140$ & 0.55 & 0.9 & 7450 & 0.16 & 0.15 & 2011-06-13\\
         & NIR  & $1020-2480$ & $1200$ & 0.55 & 0.9 & 5300 & 0.21 & 0.60 & 2011-06-13\\
GMOS     & R400 & $450-870$   & $300$  & $>1.0$ & 1.0 & 1900 & 0.14 & 1.36 & 2012A\\ 
\hline
\end{tabular}
\end{table*}

\subsubsection{\label{galextinction}Correction for galactic extinction}
J2240 (${\rm GLAT}=-54^\circ$) lies in an area of faint galactic cirrus. However, the 
particular line of sight is unaffected and has $E(B-V)=0.059$ mag and 
$R=3.3$ \citep{sfd98}. The extinction models for $R=3.3$ and the `standard' $R=3.1$ are 
indistinguishable in the wavelength range relevant for this paper \citep{fit99}. Hence we 
use the $R=3.1$ extinction curve \citep{ccm89,osf06}. Correction factors are between 1.09 
for (redshifted) [\ion{S}{II}]$\lambda$6718,33 and 1.25 for [\ion{Ne}{V}]$\lambda$3427.

\section{\label{specanalysis}2D spectroscopic analysis method}
J2240 is a dynamically complex system, and well resolved in our data. A proper description 
requires models of the NLR's 3D structure and knowledge about the radiation field, 
serving as input for photo-ionization codes. This is beyond the scope of this 
discovery paper, as we are currently lacking the X-ray data to characterize the ionizing 
spectrum, as well as IFU observations of the entire object. Nevertheless, applying 2D emission 
line diagnostics we can obtain information going beyond single global values for e.g.\ 
extinction or density. To do so, we have to bring all emission lines to a common reference 
system.

The emission lines are complex (see e.g.\ the right 
panel of Fig.~\ref{image}). We have therefore decided to use a parameter-free approach
and work with the line images directly, as compared to modeling with multiple 
Gaussians. For example, a superposition of up to four Gaussians of different widths is needed
to describe any spatial line scan through the Cloud, which fragments into several smaller 
condensations. Parts of the lines are in addition asymmetric, generally 
making a Gaussian parametrization a bad choice in this case.

\subsection{\label{o3n2ratios}Theoretical [OIII] and [NII] doublet line ratios}
The theoretical value of the [\ion{O}{III}]$\lambda$4960,5008 intensity ratio has long been 
underestimated compared to observations \citep[e.g.\ $2.89$ by][]{gmz97}. \cite{led96} 
found $3.00\pm0.08$, consistent with high-S/N measurements of other nebulae, and 
encouraged improved theoretical calculations to be carried out. 

\cite{stz00} calculate precise line ratios for [\ion{O}{III}]$\lambda4960,5008$ and
homologous cases (such as [\ion{N}{II}]$\lambda$6550,86, [\ion{Ne}{III}]$\lambda3869$, $3968$ or 
[\ion{Ne}{V}]$\lambda3347,3427$) using relativistic corrections. The intensity ratios of these 
lines are independent of temperatures and densities in typical astrophysical nebulae, as the 
two lines come from a common upper level. Only if the lines become optically thick will the 
intensity ratio change. This is unlikely, however, given their low absorption coefficients.
If the density was high enough to cause significant optical depth, collisional de-excitation 
would suppress these lines in the spectrum (Peter J.~Storey, priv.~comm.). The ratio of the 
transition probabilities between [\ion{O}{III}]$\lambda5008$ and [\ion{O}{III}]$\lambda4960$ is found to be
3.013, which translates into a flux ratio of 2.984 by multiplying with the wavelength ratio 
4960/5008. This value has been confirmed ($2.993\pm0.014$) by \cite{dpk07} based on 34 SDSS AGN 
spectra. Likewise, for [\ion{N}{II}]$\lambda6586$ and [\ion{N}{II}]$\lambda6550$ the flux ratio evaluates to 
3.05.

We use these theoretical values to deblend the H$\alpha$-[\ion{N}{II}] line complex, and to verify
our spectral analysis method with the [\ion{O}{III}] doublet.

\subsection{Re-projection of lines}
As a first step we subtract the continuum, which we linearly interpolate across
an emission line based on the adjacent $\sim100$ pixels on both sides. The resulting continuum 
is smoothed in dispersion direction with a 100 pixel wide median kernel, and subtracted 
from the original spectrum. We do not correct line fluxes for potential stellar absorption, 
as the equivalent widths are high (several 100\AAb to more than 1000\AA).

In the second step we extract images centered on each line of interest, encompassing their 
full spatial extent and 2.3\AAb in width. These images are then stretched by a factor 
of $\lambda_{\rm line}/\lambda_{{\rm H}\beta}$ to correct for the wavelength dependence of the
velocity broadening. We choose H$\beta$ as the reference frame, as it is in the middle between 
the blue [\ion{O}{II}] and red [\ion{S}{II}] lines. The re-projection onto H$\beta$ also includes a 2-fold 
binning along the dispersion direction to increase the S/N. We retain an effective spectral 
resolution of $R\sim3800$.

Thirdly, we have to register the line images such that they overlap precisely. Distortion 
correction by the XSHOOTER pipeline left no measurable spatial offsets of the continuum 
recorded in the UVB and VIS arms. Registering the lines in dispersion direction is more 
difficult, as their individual appearances are different. Fortunately, almost all lines 
exhibit traces of the ENLR with low line widths, which we use as a reference mark. Remaining 
lines are registered using other common features. In this way registration in dispersion 
direction is accurate to $\sim1$ pixel, well within the oversampled spectral resolution.

Lastly, we smooth with a 2 pixel wide Gaussian kernel, cut off at a radius of 
2 pixel. Prior to smoothing, the highest and lowest pixel in the aperture are rejected. In 
this way we suppress spurious noise features, which can get strong when calculating 
ratios with small denominators. 

Error maps from the XSHOOTER pipeline are treated analogously and fully propagated.

\begin{figure}[t]
  \includegraphics[width=1.0\hsize]{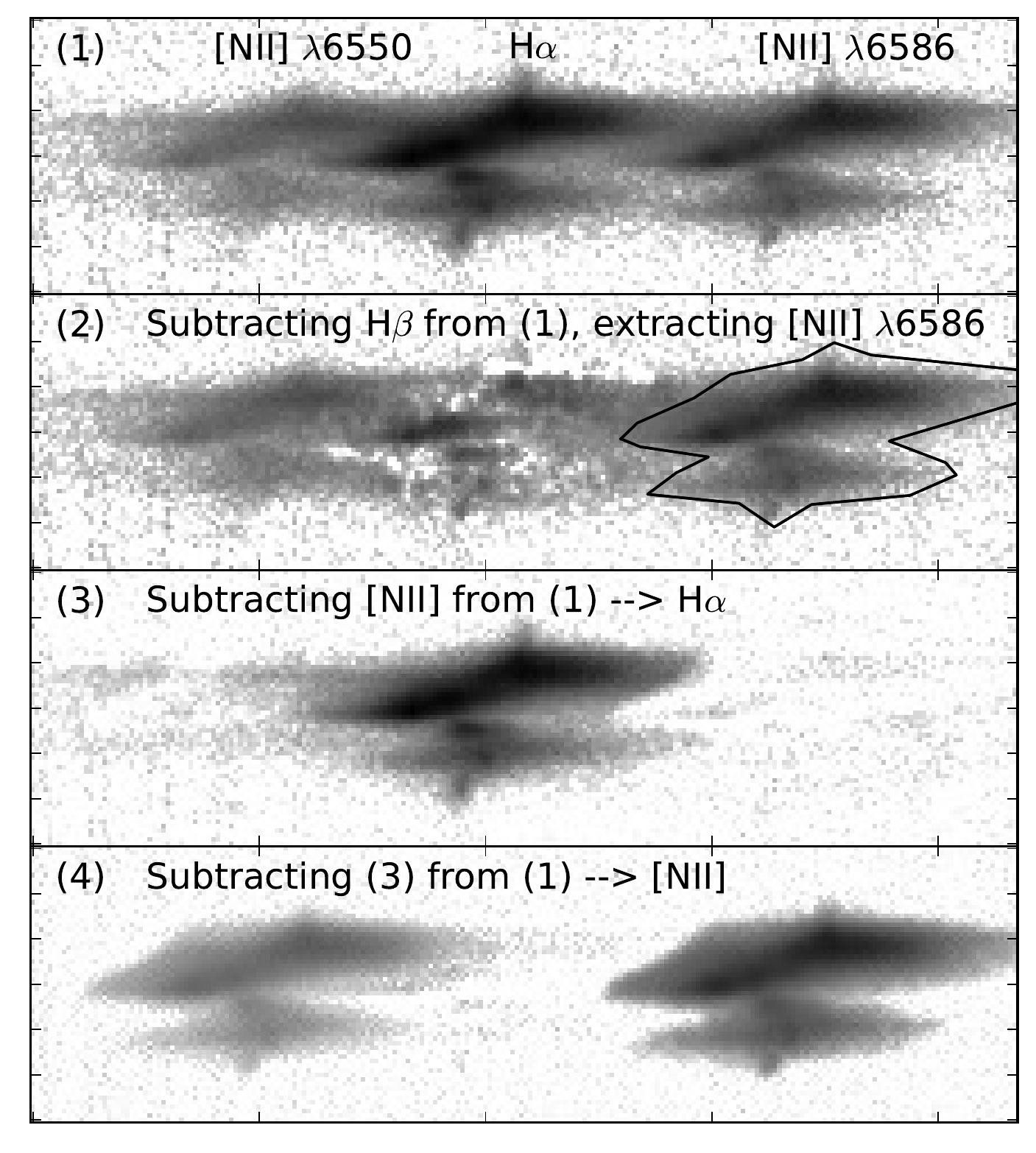}
  \caption{\label{HaNIIdeblending}{Deblending the H$\alpha$-[\ion{N}{II}] line complex (panel 1).
    Panel 2: A rescaled and smoothed version of H$\beta$ is subtracted, and [\ion{N}{II}]$\lambda6586$ 
    extracted within a closed polygon line. Panel 3: [\ion{N}{II}] gets smoothed and subtracted from the 
    original line complex, leaving H$\alpha$ isolated. Panel 4: H$\alpha$ gets smoothed and 
    subtracted from the original line complex, leaving the [\ion{N}{II}] doublet.}}
\end{figure}

\subsection{Deblending close line pairs}
Velocity broadening in J2240 leads to some overlap between H$\alpha$ and [\ion{N}{II}]$\lambda$6550,86, 
and within the [\ion{S}{II}]$\lambda$6718,33 doublet. While the bright cores of these lines are well 
separated, their fainter wings overlap. We deblend them as follows.

For the H$\alpha$-[\ion{N}{II}] complex (Fig.~\ref{HaNIIdeblending}, panel~1) we have to perform several steps.
First, we smooth H$\beta$ with a 2 pixel wide Gaussian and subtract it from H$\alpha$ after 
multiplication with 3.0. The latter is near the average Case B hydrogen ratio (3.08) in Seyfert-2 AGN 
\citep{gaf84}. In this way we remove about $90-95\%$ of the H$\alpha$ contamination in the [\ion{N}{II}] lines 
(Fig.~\ref{HaNIIdeblending}, panel~2), thus preparing a first guess of the [\ion{N}{II}] doublet. Some 
residual H$\alpha$ is still present, occurring due to variable extinction in J2240
(Sect.~\ref{extinction}).

Next we extract [\ion{N}{II}]$\lambda$6586 within a closed polygon line, then smooth and subtract it from the 
original H$\alpha$-[\ion{N}{II}] complex. We also divide it by 3.05 (Sect.~\ref{o3n2ratios}), shift it to the 
position of [\ion{N}{II}]$\lambda6550$, and subtract it once more. The [\ion{N}{II}] doublet has thus been removed and 
we have prepared a clean H$\alpha$ line (Fig.~\ref{HaNIIdeblending}, panel~3). Lastly, we smooth the 
H$\alpha$ image and subtract it from the original line complex, leaving the final [\ion{N}{II}] pair.

For [\ion{S}{II}]$\lambda6718,33$ we use a different approach, as this doublet is reasonably
separated with little overlap. We define non-overlapping closed polygonal lines of 
identical shape around both lines. Pixels within each polygon line are smoothed
with a small Gaussian kernel and subtracted, leaving the other line in the doublet 
isolated.

Note that our 2D analysis technique used in the following is confined to regions with high S/N. 
Areas where the [\ion{S}{II}] doublet or H$\alpha$ and [\ion{N}{II}] overlap are already faint and mostly excluded.

\subsection{\label{OIIIsanity}[OIII] sanity check}
As shown in Sect.~\ref{o3n2ratios}, the [\ion{O}{III}]$\lambda$4960,5008 line ratio is independent over a 
large range of temperatures and densities. Both lines are strong and uncontaminated, and thus serve as 
a good test case to validate our 2D analysis method. While a constant offset indicates a problem with 
flux calibration, systematic variations along spatial or spectral directions hint at bad 
registration of the two lines, or problems in the data processing or extinction correction.

\begin{figure}[t]
\includegraphics[width=1.0\hsize]{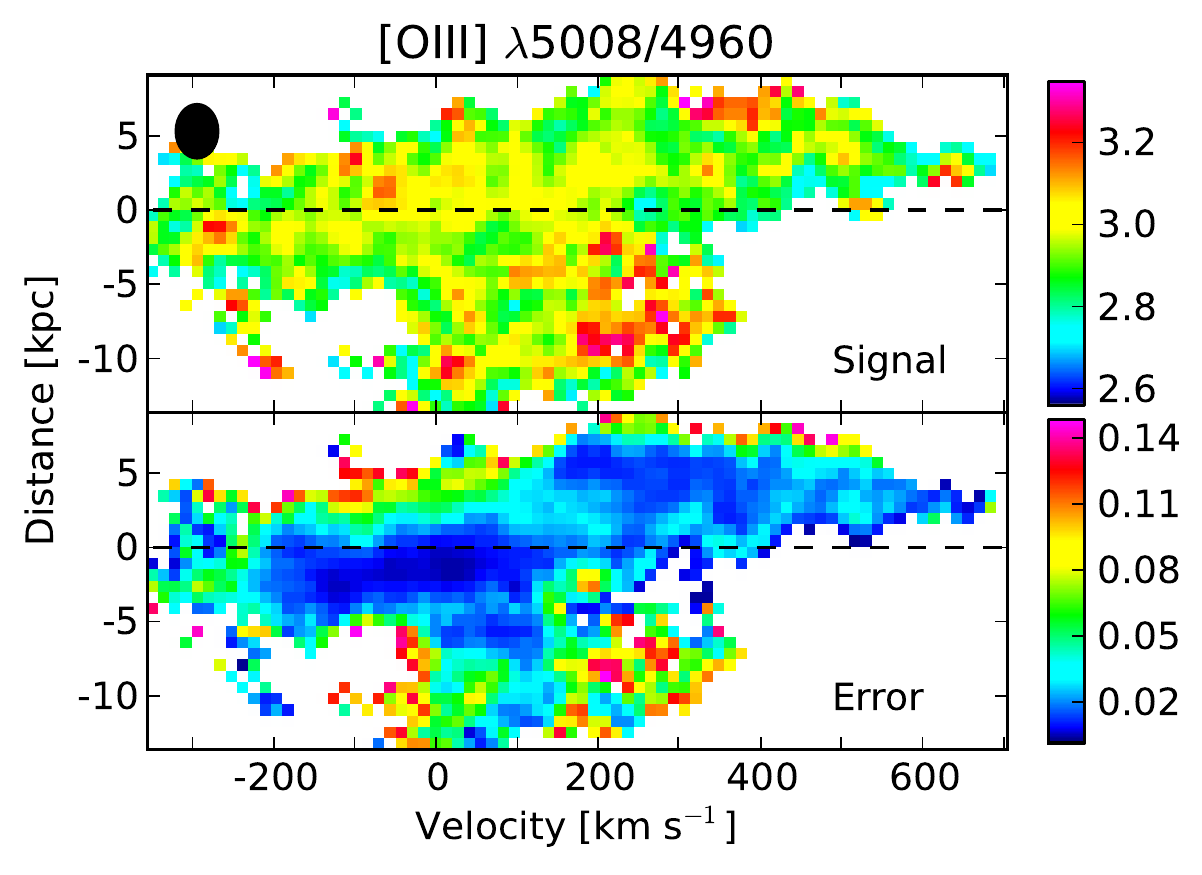}
\caption{\label{OIIIsanitycheck}The [\ion{O}{III}] $j_{\lambda5008}/j_{\lambda4960}$ ratio (theoretical 
  value: 2.98) is constant within errors. The black ellipse shows the spatial and spectral scale
  beyond which data points become statistically independent (same for all other similar plots). 
  The dashed line indicates the spatial location of highest continuum flux (coincident with the 
  nucleus). The origin of the velocity axis coincides with the location of highest electron 
  density (see Sect.~\ref{temden}).}
\end{figure}

Figure~\ref{OIIIsanitycheck} shows that in our data the [\ion{O}{III}] ratio is constant within errors, 
thus validating our 2D analysis method. Utilizing the area where $j_{\lambda5008}/j_{\lambda4960}$ 
uncertainty is less than 0.03, we initially estimated a line ratio of $3.169\pm0.012$ 
(including full extinction correction, see Sect.~\ref{extinction}). This is inconsistent with 
the theoretical value, and amongst the highest values measured by \cite{dpk07}. Retracing our 
processing steps we identified an unfortunate coincidence causing this discrepancy. The 
[\ion{O}{III}]$\lambda4960$ line is redshifted to $6577$\AA, just redwards of H$\alpha$. Being a white 
dwarf, the spectrophotometric standard Feige 110 shows a $\sim20$\AAb wide H$\alpha$ absorption 
line in the XSHOOTER spectrum. The reference flux values tabulated by \cite{oke90}, however, are
based on a low resolution spectrum ($\sim5$\AAb pixel$^{-1}$), resulting in a measured line 
width of $\sim40$\AA. As the [\ion{O}{III}] line is narrow ($1.5$\AA) compared to the absorption 
line, we can easily determine a correction factor of $1.075\pm0.015$. The [\ion{O}{III}] intensity ratio 
then becomes $2.950\pm0.015$. The same effect is present in our FORS2 data (for which a different 
white dwarf served as flux calibrator), yielding a corrected ratio of $3.01\pm0.04$. No other
emission lines are affected.

\section{Results from the spectral analysis}
We use bright emission lines to determine extinction, electron density and temperature, abundance 
and ionization. These diagnostics are well understood for various ionizing spectra.
However, as the X-ray spectrum of J2240 is not available, uncertainties do exist in some of the 
parameters derived. We address this where applicable.

\subsection{\label{extinction}Internal extinction}
\subsubsection{\label{iterationmethod}Method}
The extinction of a line with intrinsic intensity $I_{\lambda0}$ is given as 
$I_{\lambda}= I_{\lambda0}\,{\rm e}^{-C\,f(\lambda)}$, where $C$ is a constant, and $f(\lambda)$ 
parametrizes the extinction curve of the interstellar medium, and depends on the chemical and 
physical properties of the dust grains. Using the observed and intrinsic (before attenuation by 
dust) intensity ratio of two emission lines, the color excess is
\begin{equation}
E(B-V) = -2.5\,{\rm log} \left(\frac{I_\lambda/I_\nu}{I_{\lambda0}/I_{\nu0}}\right) 
\frac{f(B)-f(V)}{f(\lambda)-f(\nu)}\;. 
\end{equation}

\begin{figure*}[t]
\includegraphics[width=1.0\hsize]{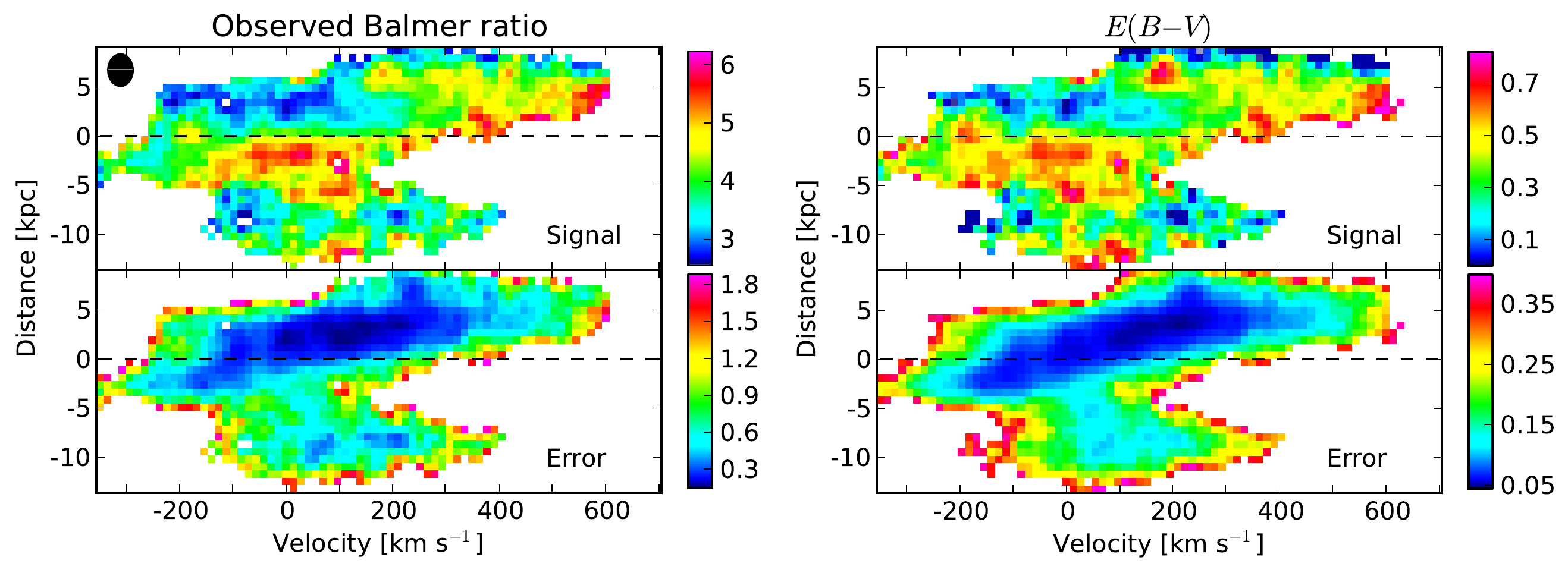}
\caption{\label{extinctionall}{Observed H$\alpha$/H$\beta$ ratio (left), and the reddening inferred 
    (right).}}
\end{figure*}

We use the H$\alpha$/H$\beta$ ratio and the standard $R=3.1$ model \citep{ccm89} to calculate
the reddening in J2240. This requires knowledge of the intrinsic value of H$\alpha$/H$\beta$,
which equals $\sim2.87$ for Case B conditions in a typical \ion{H}{II} region. In the harder radiation field 
of an AGN collisional excitation of H$\alpha$ becomes important, increasing the ratio to $\sim3.08$. 
However, it depends strongly on $L_X/L_{\rm opt}$ and the slope of the X-ray spectrum, as well as on 
metallicity \citep{gaf84}. The intrinsic line ratio can get as low as $2.6$ for high 
metallicities, or as high as 3.4 for low metallicities.

We take the metallicity into account in an explicit manner. Starting with a fixed intrinsic Balmer
line ratio of 3.0, we compute an initial metallicity estimate following \cite{ssc98}.
Once a first guess for the metallicity is available, we update the intrinsic Balmer line ratio using an 
interpolation function to the values given in \cite{gaf84}. Pixel values now deviate individually from the 
initial guess of $3.0$. With the improved extinction map we calculate a new abundance map. Convergence 
is achieved rapidly (less than 1\% relative change for most pixels) after the first iteration, thus we 
stop after a second iteration. Different starting point values of 2.7 and 3.2 for the intrinsic Balmer 
ratio converge to the same solution.

\subsubsection{The reddening map}
The resulting reddening map is shown in the right panel of Fig.~\ref{extinctionall}. It is 
based on the Seyfert 1 model X-ray spectrum of \cite{gaf84}. Adopting their $\nu^{-1}$ power law 
continuum (dashed line in their Fig.~1), $E(B-V)$ is systematically reduced by 0.16 mag, but the 
structures seen remain unchanged. We work with the higher extinction model for the rest of this 
paper, commenting on the effects of a different X-ray spectrum where applicable.

The dust in J2240 is distributed unevenly. The area between the nucleus and the Cloud is highly reddened
with $E(B-V)=0.5-0.7$ mag. It is independent of radial velocity, and likely caused by a large cloud of 
foreground dust in J2240. On the opposite side of the nucleus (above the dashed line in 
Fig.~\ref{extinctionall}), we find that reddening grows as a function of radial velocity (blue-shifted 
components are less reddened than redshifted ones). In this area the dust appears to be evenly embedded 
within the NLR \citep[where it can survive, see][]{lad93}, causing $E(B-V)$ to increase from $0.05$ to 
$\sim0.7$ mag as we look deeper into J2240. Other Seyfert-2 galaxies show similar (global) extinction 
values of $E(B-V)=0.1-0.6$ for their NLRs \citep{hfs97,rhl05}.

\subsubsection{Error analysis}
The intrinsic H$\alpha$/H$\beta$ ratio is affected by collisional excitation and metallicity. The latter
was taken into account in Sect.~\ref{iterationmethod}. As an additional test, we artificially 
increase the metallicity by $\Delta Z=+0.2\,Z_\odot$, which reduces $E(B-V)$ by $0.031\pm0.005$ mag only
(see also Sect.~\ref{abundance}). The largest 
uncertainty comes from the unknown X-ray properties of the AGN. Using the values obtained by \cite{gaf84} 
for a Seyfert-1 X-ray spectrum, we get a mean intrinsic Balmer ratio of $2.75$ with small spatial 
variations of $0.02-0.05$ mag. This is below the commonly adopted value of 3.08 for Seyfert-2 
galaxies, and may be real, or the result of our ignorance about the true X-ray properties. The 
$\nu^{-1}$ power law continuum from \cite{gaf84} yields $3.11\pm0.04$ for the intrinsic line ratio, 
reducing $E(B-V)$ by 0.12 mag, which is small compared to the reddening values up to 0.7 mag. We thus
conclude that the observed variations in H$\alpha$/H$\beta$ are indeed due to dust, and not caused by 
spatial variations in abundance (see also Sect.~\ref{abundance}) nor the ionizing spectrum.

Another source of uncertainty is the accuracy with which the Balmer lines have been registered. A Gaussian 
fit to the continuum profiles near H$\alpha$ and H$\beta$ yields an uncertainty of 0.15 pixels in the spatial 
direction. Being conservative, we shift H$\alpha$ with respect to H$\beta$ by 0.3 pixel in both spatial 
directions. This introduces a maximum gradual change in $E(B-V)$ of $0.32$ mag between $+7$ kpc and $-3$ 
kpc, about half the observed amplitude of 0.7 mag. Offsetting H$\alpha$ by $\pm1$ pixel along the spectral 
direction results in a 0.18 mag change between $-200$ and $+100$ km s$^{-1}$, and no change elsewhere.
This highlights the importance of an accurate registration of the Balmer lines. Assuming a worst-case
scenario, we found that the observed extinction features (in particular the large dust cloud) prevail.
We can thus realistically estimate a maximum registration error in $E(B-V)$ of 0.15 mag at any given 
position.

Alternatively, $E(B-V)$ can be obtained using Pa$\alpha$/Pa$\beta$ or other Balmer ratios. However, we 
cannot use the Paschen series as redshifted Pa$\alpha$ is beyond XSHOOTER's NIR arm, and the core of Pa$\beta$ 
is masked by airglow. Using the H$\gamma$/H$\beta$ ratio has the advantage of not being affected by 
collisional excitation (and thus the AGN spectrum). Unfortunately, redshifted H$\gamma$ falls in the 
area where XSHOOTER's optical dichroic distinguishes between the UVB and VIS arms, significantly 
reducing the flux, and making flux calibration unreliable. The S/N of the H$\delta$ line is too low to 
perform a 2D analysis. Integrating over the whole lines, both H$\delta$ and H$\gamma$ favor an average 
reddening of $E(B-V)=0.52-0.61$ mag, $0.2-0.3$ mag higher than predicted by H$\alpha$. This can be 
explained if stellar absorption, for which we do not correct the continuum model, reduces the flux of 
the fainter Balmer lines by $15-20$\%.
 
Another consistency check is the extinction corrected [\ion{O}{III}] ratio presented in Sect.~\ref{OIIIsanity}. 
The extinction map shows a large area below the nucleus of $E(B-V)=0.5-0.7$ mag, embedded in lower 
values of $0.1-0.2$ mag. If this differential extinction was artificial or the [\ion{O}{III}] ratio was left
uncorrected, then $j_{\lambda5008}/j_{\lambda4960}$ increases by 0.07 for the highly reddened area (with 
respect to its surroundings). Such an increase is not found in the data, thus confirming this highly 
reddened area. This test is not very sensitive though, as the flux ratio of the relatively close
[\ion{O}{III}] doublet is weakly susceptible to differential reddening effects.

\subsubsection{\label{globalextinct}Testing different dust models}
The composition of dust in the harsh environment of an AGN is still a matter of debate, 
in particular in view of the unified dusty torus model \citep[e.g.][and references 
therein]{rhl05}. While dust cannot survive inside the broad line region \citep{lad93}, 
larger grains prevail in the circum-nuclear region further out, leading to flat extinction 
curves \citep[gray dust, see e.g.][]{mms01}. At larger distances from the central AGN engine, smaller 
dust grains can survive or are replenished by the partial destruction of larger grains. However, very 
small grains are disintegrated quickly even at kpc scales \citep{voi92}. A large number of distant 
AGN and narrow-line quasars show extinction compatible with a dust composition like the one observed 
in the Small Magellanic Cloud \citep[SMC;][]{hsh04,wil05}. Even steeper extinction curves in 
AGN have been found by \cite{ckb01,ckt02} for NGC 3227 and Ark 564 at wavelengths below 4000\AA.

Applying the SMC dust model instead of $R=3.1$ causes no significant change to $E(B-V)$ 
($0.04$ mag) as compared to the global variations present in J2240. This is expected since
these models differ only slightly at the H$\alpha$ and H$\beta$ wavelengths
\citep[][]{ccm89,fit99}. Both foreground dust (close to J2240) as well as dust embedded in 
the NLR exist in J2240. Observing lines further into the UV or the IR will allow us to step beyond 
the standard $R=3.1$ model and constrain the actual physical properties of the dust in J2240, 
facilitated by the bright emission lines.

\begin{table}[t]
\caption{Electron temperatures}
\label{electrontemperature}
\begin{tabular}{lrr}
\noalign{\smallskip}
\hline
\hline
\noalign{\smallskip}
Line & $T_{\rm disk}$ [$10^3\,$K] & $T_{\rm cloud}$ [$10^3\,$K] \\
\hline
\noalign{\smallskip}
[\ion{O}{III}] & $17.1_{-0.7}^{+0.8}$ & $14.7_{-2.0}^{+4.8}$ \\
\noalign{\smallskip}
${\rm [\ion{Ne}{III}]}$ & $18.1_{-2.4}^{+3.2}$ & $15.1_{-4.1}^{+\infty}$ \\
\noalign{\smallskip}
[\ion{O}{II}] & $13.0_{-1.3}^{+1.4}$ & \nodata \\
\hline
\end{tabular}
\end{table}

\begin{figure*}[t]
  \includegraphics[width=1.0\hsize]{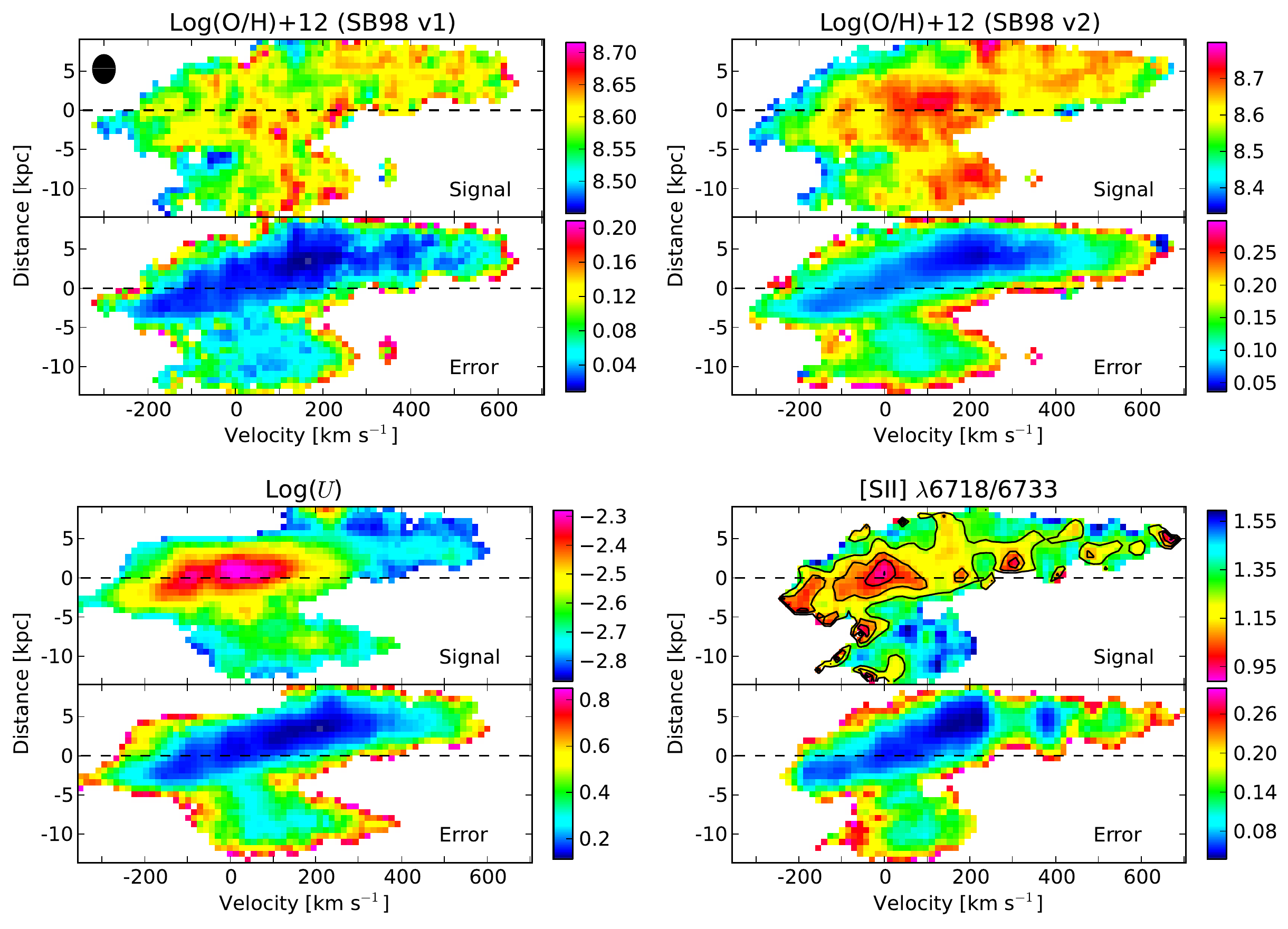}
  \caption{\label{abu_ion_dens}Top row: Oxygen abundance, estimated using the two strong line 
    diagnostics of \cite{ssc98}. Bottom left: Ionization parameter.
    Bottom right: [\ion{S}{II}] line ratio (map) and electron density (contours) for $T_{\rm e}=13000$ K. 
    Contour levels are drawn at $n_{\rm e} = 150, 300, 450, 600$ cm$^{-3}$.}
\end{figure*}

\subsection{\label{temden}Electron temperature and density}
The electron temperature, $T_{\rm e}$, for the medium ionization zone can be obtained from [\ion{O}{III}],
utilizing 
\begin{equation}
\frac{j_{\lambda4960}+j_{\lambda5008}}{j_{\lambda4364}} = 
\frac{7.9 \, {\rm exp} (3.29\times 10^4 / T_{\rm e})}{1+4.5 \times 10^{-4}\,n_{\rm e} / T_{\rm e}^{0.5}}
\end{equation}
\citep{osf06}. Similar relations exist for [\ion{Ne}{III}] ($(j_{\lambda3869}+j_{\lambda3968})/j_{\lambda3343}$) 
and [\ion{N}{II}] ($(j_{\lambda6548}+j_{\lambda6583})/j_{\lambda5755}$). For electron densities 
$n_e\lesssim10^5$ cm$^{-3}$ these temperature probes become independent of $n_{\rm e}$. 
As the redshifted (and weak) [\ion{N}{II}]$\lambda5755$ line is lost in the atmospheric O$_2$ absorption feature 
at $7630$\AA we cannot use [\ion{N}{II}] to probe the lower ionization regions. Instead, we use [\ion{O}{II}] 
$j_{\lambda3727}/j_{\lambda7321,32}$ and eq.~(21) from \cite{pei67}, together with an average electron 
density of 200 cm$^{-3}$ for the Disk.

Due to the weak auroral lines we cannot reconstruct 2D temperature maps, yet we can still measure 
global values. Using [\ion{O}{III}] we find 17100 K and 14700 K for the Disk and the Cloud, respectively 
(Table~\ref{electrontemperature}). [\ion{Ne}{III}] results in 18100 K for the Disk, whereas the 
Cloud is undetected in [\ion{Ne}{III}]$\lambda3343$ and only a lower limit of 11000 K can thus be 
obtained. Errors account for the significant blending of [\ion{Ne}{III}]$\lambda3343$ with the much 
brighter [\ion{Ne}{V}]$\lambda3347$. For the lower ionization zone we find 13000 K from [\ion{O}{II}]. 

The [\ion{S}{II}] $j_{\lambda6718}/j_{\lambda6733}$ ratio depends on $n_{\rm e}$ and 
$T_{\rm e}$. We use $T_{\rm e}=13000$ K as obtained from [\ion{O}{II}] for the low ionization zone. 
Results are shown in the lower right panel of Fig.~\ref{abu_ion_dens}. The Disk has significantly 
higher density than the rest, starting from $150-200$ cm$^{-3}$ and peaking in 
a dynamically cool spot 1 kpc north-east of the nucleus where $n_{\rm e}$ reaches 650 cm$^{-3}$.
Decreasing (increasing) $T_{\rm e}$ by 5000~K rises (lowers) the peak density by 130 (80) 
cm$^{-3}$, which is significantly less than the density variations measured.
These features are thus real, and not a consequence of our simple assumption of constant temperature.

The Cloud is in the low density limit of the [\ion{S}{II}] probe, thus only an upper limit of 
$n_{\rm e}<50$ cm$^{-3}$ can be inferred. The area north-east of the nucleus, where J2240 is brightest 
in [\ion{S}{II}], has lower density than the Disk ($80-150$ cm$^{-3}$). 

\subsection{\label{abundance}Abundance and ionization}
Abundances for \ion{H}{II} regions are often obtained with the strong-line method \citep{ked02}. A more accurately 
approach is the direct method \citep[e.g.\ 
implemented in the {\tt nebular/ionic} task in {\tt IRAF} based on the five level atomic model by][]{ddh87}. 
Abundance determinations for NLRs are notoriously more difficult. \cite{ssc98} have developed an empirical 
strong-line method based on NLRs with embedded \ion{H}{II} regions. The underlying assumption is that both the 
\ion{H}{II} regions and the NLR share the same metallicity, thus a calibration for NLRs without discernable
\ion{H}{II} regions can be inferred. \cite{ssc98} offer two methods, one based on [\ion{N}{II}]/H$\alpha$ and [\ion{O}{III}]/H$\beta$, 
and one on log$({\rm [\ion{O}{II}]/[\ion{O}{III}]})$ and log$({\rm [\ion{N}{II}]/H\alpha})$. The latter depends on extinction 
correction as [\ion{O}{II}] and [\ion{O}{III}] are not neighboring lines. In this work we also included their correction 
term for electron density. 

Both methods yield similar results of $Z=(0.45-0.55)Z_\odot$, the second one with slightly larger values 
of $\Delta{\rm log(O/H)}=0.03$ (Fig.~\ref{abu_ion_dens}). Abundances are enhanced by about 0.1 towards the 
nucleus and in the Cloud, and appear to increase systematically with growing radial velocity. Such sub-solar 
metallicities are uncommon in NLRs, as has been shown by \cite{ghk06}. Their study of 23000 Seyfert-2 
galaxies in SDSS yielded only $\sim$40 galaxies of evidently low metallicity. Using their prescription we 
obtain average metallicities of $0.9 Z_\odot$ (from [\ion{N}{II}]/[\ion{O}{II}] and [\ion{O}{III}]/[\ion{O}{II}]), and $\sim2.0 Z_\odot$ 
(from [\ion{N}{II}]/H$\alpha$ and [\ion{O}{III}]/H$\beta$). 

A large uncertainty in the abundance determination stems from the assumptions made about the ionizing 
spectrum. \cite{lgb11} report that changes to the slope and shape of the spectrum can change metallicities 
from sub-solar level to several times the solar value. As we do not know the intrinsic X-ray spectrum of the 
AGN in J2240, the true absolute value of the abundance remains unknown. The variations measured in the abundance 
map may be real, but can be altered by local variations of the X-ray spectrum filtered by the 
inhomogeneous ISM. We therefore do not consider abundances further in this work, leaving this task to future 
photo-ionization modeling once X-ray data become available. Also, note that increasing the metallicity from 
$0.5\,Z_\odot$ to several $Z_\odot$ changes the intrinsic H$\alpha$/H$\beta$ Balmer ratio from 2.75 to about 
$2.60-2.65$, increasing $E(B-V)$ only marginally.

Once the metallicity is known, we can derive the ionization parameter $U$ (Fig.~\ref{abu_ion_dens}, lower left) 
from [\ion{O}{III}]/[\ion{O}{II}] \citep{ked02}. Applying the best-fit relation from \cite{pra90} we find similar structures
and values (not shown). Using [\ion{O}{III}]/[\ion{O}{II}] requires $n_e<10^3$ cm$^{-3}$ \citep{kos97}, since mixed-in higher 
densities will lead to an over-estimation of $U$. This condition is well met (Fig.~\ref{abu_ion_dens}, 
lower right).

We find two strong peaks in the ionization map located $0.0-2.0$ kpc north-east of the nucleus 
(above the dashed line in Fig.~\ref{abu_ion_dens}), and spread over a radial 
velocity range of 210 km s$^{-1}$. The spatial offset is coincident with that measured for
the density peak, and can either indicate the location of a second, deeply buried AGN, or simply 
be the result of a shock or interaction with a jet. Two weaker peaks are found 4.5 and 8 kpc south-west 
of the nucleus (below the dashed line), significant on the 3 and 2$\sigma$ level, respectively. The first
is located between the nucleus and the Cloud in the area with highest dust extinction, whereas the second 
is centered in the Cloud and redshifted by 110 km s$^{-1}$ with respect to the highest ionization peak.

The error map for $U$ takes into account measurement errors, but not the uncertainty in 
metallicity. To test the latter, we lower the metallicity by $\Delta Z=0.1\,Z_\odot$, which corresponds 
to going from high to low metallicity areas. ${\rm Log}\,U$ then decreases by maximally 0.12, much less 
than the absolute variations observed ($\Delta ({\rm log\,U})=0.85$), and less than the direct measurement 
errors. Note that while the mentioned variations in $U$ are real, its absolute value should not be taken 
at face value, as the absolute metallicity is unknown.

\subsection{\label{continuum}Continuum}
J2240 exhibits a relatively flat continuum with the brightest parts at $4500$\AA$-7000$\AAb
rest-frame wavelengths (Fig.~\ref{emissionlines}). With an extent of about 8 kpc in the
2D spectra, continuum radiation emerges from a more compact region than the line emission.
A significant fraction of the continuum is stellar, as we clearly see the \ion{Ca}{II}~H absorption 
line at 3934\AA. \ion{Ca}{II}~K at 3969\AAb is superimposed by [\ion{Ne}{III}] and H$\epsilon$ emission. 
The 4000\AAb break is hardly visible in the data.

\begin{figure}[t]
\includegraphics[width=1.0\hsize]{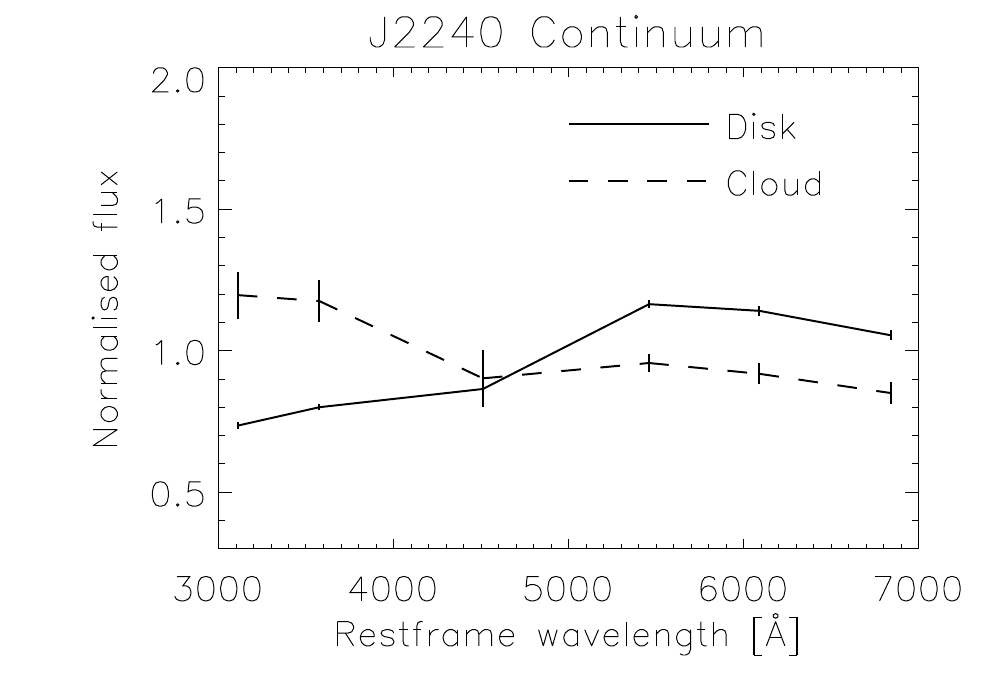}
\caption{\label{continuumspectra}{By integrating the faint continuum over large wavelength ranges, 
    excluding emission lines, we probe the photometric properties of the continuum. The Cloud is bluer
    than the Disk.}}
\end{figure}

No continuum can be seen in the 2D spectra at the position of the Cloud. However, we can 
still test for the presence of continuum flux by integrating over all wavelengths excluding
emission lines. The S/N obtained is sufficient to reconstruct broadband continuum 
SEDs for the Disk and the Cloud (Fig.~\ref{continuumspectra}). 
While their colors are similar above 5500\AA, the Cloud is bluer than the Disk at shorter 
wavelengths. Both a nebular continuum and a different stellar population can cause this. 

We find a symmetric continuum profile within $5$ kpc of either side of the nucleus 
(Fig.~\ref{lineprofile}). In the north-east the profile is nearly exponential with a Sersic index of 
$0.85\pm0.04$ and a disk scale length of $3.0\pm0.1$ kpc. In the south-west we see a long tail 
through the Cloud (the bump at $-10$ kpc in Fig.~\ref{lineprofile}), traceable over 18 kpc.

\subsection{\label{velocities}Velocities and an ENLR}
\begin{figure*}[t]
\includegraphics[width=1.0\hsize]{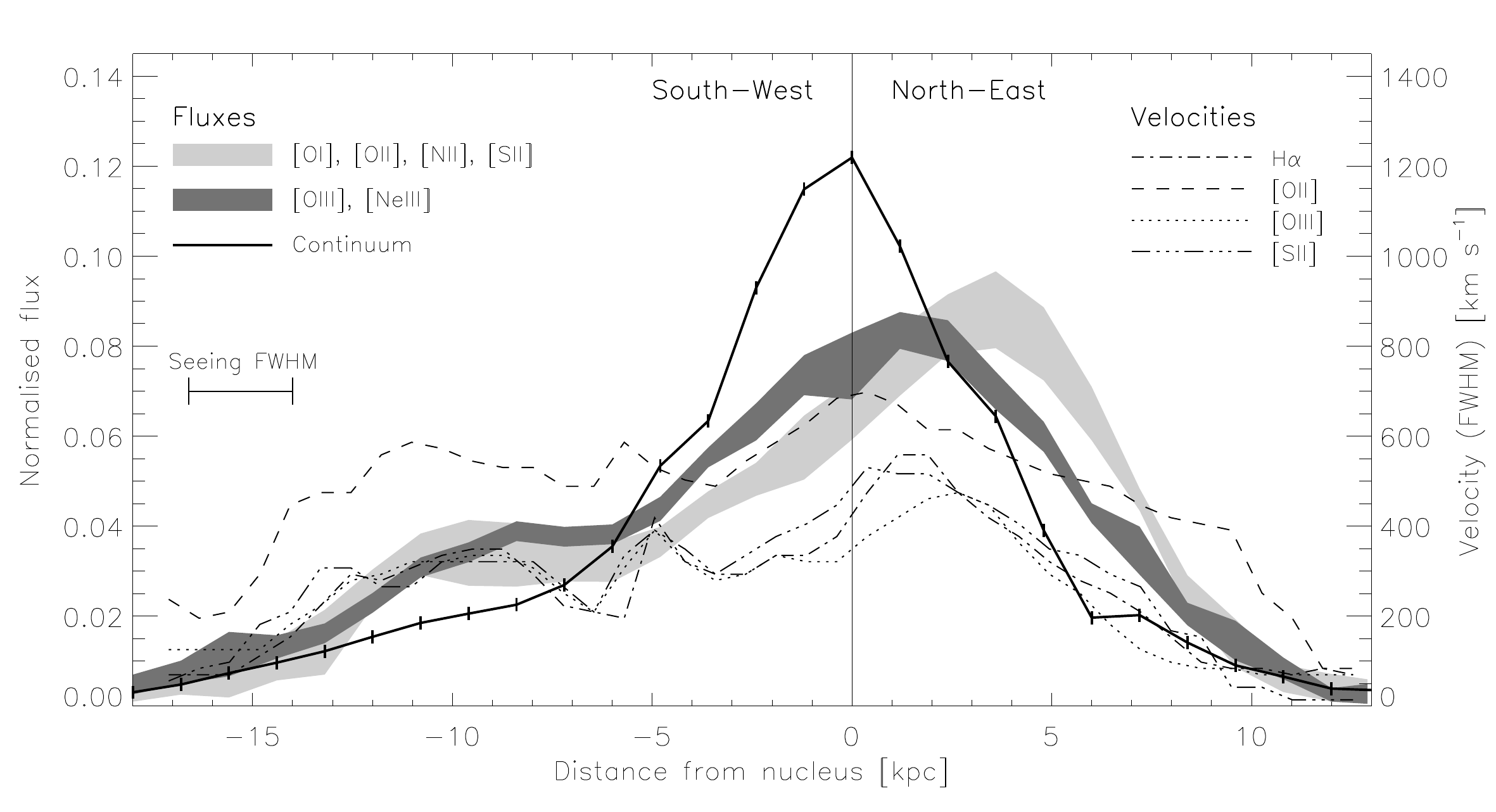}
\caption{\label{lineprofile}{Total flux as a function of slit position, 
    normalized to unity, for a selection of emission lines and the continuum (shaded areas and 
    thick line, respectively). The Cloud forms the bump at -10 kpc. Low ionization lines ([\ion{O}{I}], 
    [\ion{O}{II}], [\ion{N}{II}], [\ion{S}{II}]) peak at 3.7 kpc from the nucleus, whereas medium ionization lines 
    ([\ion{O}{III}], [\ion{Ne}{III}]) reach a maximum at 1 kpc. H$\alpha$ runs between the two. The widths of
    the various lines are over-plotted (thin lines). While H$\alpha$, [\ion{O}{III}] and 
    [\ion{S}{II}]$\lambda6718,33$ have similar FWHMs, [\ion{O}{II}] is significantly broader. Error bars 
    (not shown) for the velocity profiles are $10-15$ km s$^{-1}$.}}
\end{figure*}

The velocity FWHMs (parameter-free, directly measured in the line images) as a function of slit position 
for selected lines are displayed in Fig.~\ref{lineprofile}, together with their intensity profiles. While 
the latter for low and medium ionization lines are well distinguished (bright and dark gray areas) within 
8 kpc of the nucleus, the velocity profiles are similar. The notable exception is [\ion{O}{II}] whose FWHM is 
consistently larger by $222\pm15$ km s$^{-1}$, but otherwise has the same shape. This is a result of the 
absence of compact bright cores in [\ion{O}{II}] which are found in e.g.\ H$\alpha$, [\ion{O}{III}] and [\ion{S}{II}] (see 
Fig.~\ref{lines_all} in the Appendix for the log-scaled 2D line images).

Defining the maximum velocity as the point where the line profile drops to 20\% of the maximum emission, 
we find $v_{\rm max}=(0.98\pm0.05)\,\times\,{\rm FWHM}$ averaged over H$\alpha$, [\ion{O}{III}] and [\ion{S}{II}]. We find
little variation ($\pm0.14$) along the slit. Highest velocities are blue-shifted 1200 km s$^{-1}$. 

The [\ion{O}{II}] velocity profile is remarkably flat between $-14$ kpc and $+8$ kpc, with line widths of $534\pm69$ 
km s$^{-1}$. H$\alpha$, [\ion{O}{III}] and [\ion{S}{II}] vary between 200 and 550 km s$^{-1}$, and drop to less than 100 km 
s$^{-1}$ for the largest nuclear distances. J2240 is embedded in an ENLR that co-rotates with the disk (albeit 
slower, the north-east part is redshifted by 195 km s$^{-1}$ with respect to south-west). The ENLR's low 
luminosity, low line width, high excitation state (${\rm log([\ion{O}{III}]/H}\beta)=0.86-0.98$) and rotation are 
typical \citep{upa87,dur89}.
 
\begin{figure}[t]
\includegraphics[width=1.0\hsize]{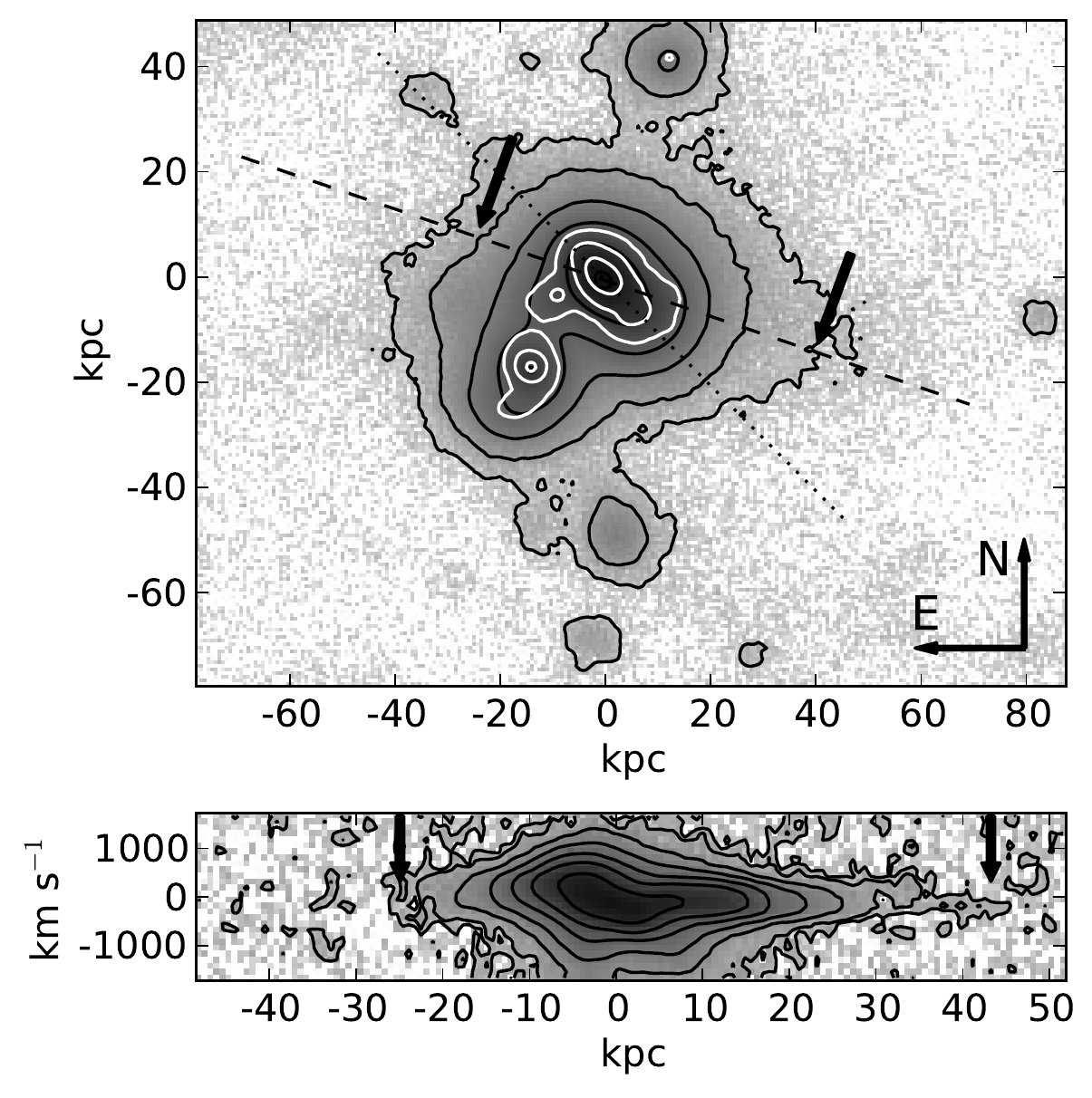}
\caption{\label{halo}Deep GMOS $r$-band image (top) of J2240 revealing the extended halo. The 
  dashed line indicates the position angle of the spectrum (bottom, showing the [\ion{O}{III}] line), 
  the two arrows out to which radius we detect [\ion{O}{III}]. Black contour lines are logarithmically 
  spaced. Contours (white) of the CFHT image (Fig.~\ref{image}) and the XSHOOTER slit (dotted line) are 
  overplotted for reference.}
\end{figure}

Deep GMOS imaging reveals a halo stretching 60 kpc from the nucleus (Fig.~\ref{halo}, 
top panel). To distinguish between a stellar tidal stream and ionized gas we use a 3h GMOS long-slit 
spectrum through a 1\myarcsec5 slit, aligned along the major halo axis (at a position angle of 71 degrees). 
Unresolved [\ion{O}{III}]$\lambda5008$ is detected out to 42 kpc (Fig.~\ref{halo}, bottom panel), thus an additional 
underlying stellar population can currently not be ruled out. The ENLR has a minimum diameter of $70$ kpc.

\subsection{\label{SBorAGN}Starburst or an AGN?}
So far we have anticipated that the NLR is powered by an AGN. Observational evidence is 
presented in the following.

Hypothetically assuming that all line emission is powered by stars, we can estimate the necessary star 
formation rate using the extrapolated total line fluxes from Table~\ref{lineflux} and the scaling relations 
from \cite{ken98}. The result equals $290\pm60\, M_\odot\, {\rm yr^{-1}}$ needed to explain the
H$\alpha$ flux, 20 times higher than that in an average star-forming GP galaxy \citep{css09}.


An AGN can be distinguished from star formation due to its harder radiation field changing the line ratios. 
\cite{bpt81} present various tools for this purpose. Their BPT diagrams have been further developed by
\cite{veo87}, \cite{kds01}, \cite{kht03} and \cite{kgk06}. For J2240 the S/N is good enough to run 
this analysis on the basis of individual pixels. While the precise locations of the dividing lines 
between SF, AGN, LINERs and composite objects in the BPT plots are still under some debate \citep{css10}, 
the classification of J2240 as AGN is beyond doubt (Fig.~\ref{linediagnostics}). [\ion{O}{I}]$\lambda6302,6366$ 
emission across J2240 shows that shock/ionization fronts are present on a global 
level, emphasizing a powerful AGN.

When fitting the continuum across emission lines, we have ignored stellar absorption as the emission 
lines are much brighter than the continuum. If stellar absorption was present, H$\alpha$ and 
H$\beta$ fluxes are underestimated because of our negligence, driving the data points in the 
BPT diagrams away from star formation. Assuming a worst-case scenario with (uncorrected) 100\% stellar 
absorption, BPT values are still far outside the star-forming area (see arrows in Fig.~\ref{linediagnostics}).

While star formation appears to be negligible in J2240, \cite{bjk06b} warn that \ion{H}{II} regions can dominate 
the [\ion{O}{III}] flux at larger nuclear distances. \cite{lev07} and \cite{kht03} also show that intense star 
formation is common in active galaxies. Indeed, a significant star burst can be buried by the [\ion{O}{III}] 
emission in J2240. However, while we do observe a decrease in log([\ion{O}{III}]/H$\beta$) with increasing radius 
(middle left panel in Fig.~\ref{linediagnostics}), the AGN characteristics remain well preserved. In 
addition, if the NLR is due to a massive outflow of hot gas, the latter may have disrupted star formation 
\citep[for an example see][]{cmm12}.

\section{\label{gbsurvey}A sample of GBs}
\subsection{Selection and verification}
J2240 has similar colors as typical GP galaxies. Applying the GP filter of \cite{css09}
to SDSS-DR8 we recover J2240 once we drop their maximum Petrosian radius of 2\myarcsec0. Instead, 
we require a minimum radius of 2\myarcsec0 and modified color cuts to select objects with
particularly strong [\ion{O}{III}] fluxes ($g-r>1.0$). $95\%$ of the objects found are 
spurious or have corrupted photometry (the latter in particular near ${\rm RA}=355\pm5$ and 
${\rm DEC}=60\pm10$; 
see examples in Fig.~\ref{gbsample-figs}). If in doubt we downloaded the original SDSS FITS 
images and created the color poststamps ourselves. In 100\% of these cases the galaxies turn 
out to have normal colors or be artifacts.

Our SQL query for objects with $0.12<z<0.35$ (where [\ion{O}{III}] falls into the $r$-band) is 
shown in Appendix \ref{gbsample}. Shifting each bandpass by one filter towards redder 
wavelengths (e.g.\ replacing $g-r$ with $r-i$, and leaving out the constraints requiring 
filters redder than $z$-band), we also select possible candidates at $0.39<z<0.69$.
Contamination by lower redshift star-forming galaxies with strong H$\alpha$ emission 
is expected in this higher redshift bin.

Final manual selection results in a sample of 29 candidates (Table~\ref{gbsample-table}, and 
Fig.~\ref{gbsample-figs}). Redshifts and [\ion{O}{III}]/H$\beta$ ratios have been measured with GMOS.
All of our low redshift candidates observed show high [\ion{O}{III}]/H$\beta$ ratios, confirming their AGN character. 
Broad line components are absent but not ruled out due to shallow data. Seven galaxies have a Petrosian 
radius between 2\myarcsec4 and 2\myarcsec6 and are good examples of GBs with particularly large 
NLRs. J2240 (listed as \#016) is larger than any of the other galaxies (Petrosian radius of 3\myarcsec5). 
\#004 is noteworthy, as it shows a large extension which indicates ejected matter or tidal interaction.
Several GBs exhibit close companions of unknown redshift, indicating that encounters may play a role 
in the formation of GBs.

In the higher redshift sample we find 12 objects, of which 7 have spectra taken. Five galaxies have genuine 
AGN spectra, whereas 2 are star-forming galaxies at lower redshifts for which H$\alpha$ was mistaken 
as [\ion{O}{III}] by our SQL filter. All objects covered by the VLA FIRST catalog \citep{wbh97} are either 
non-detected or radio-quiet.

Apart from \#015 all spectroscopically confirmed GBs have [\ion{O}{III}] lines extending over $15-20$ 
kpc or more (Fig.~\ref{gbsample_OIII}). Given integration times of only 300s, the real extent of 
the NLRs is likely to be larger. In addition, the limited availability of guide stars is restricting 
the slit position angle, which consequently cannot be well aligned with the target's major axis in 
a majority of cases.

\subsection{\label{midir}Mid-IR properties}
Type-2 AGN are dusty objects with significant optical extinction. The WISE \citep{wem10} mid-infrared fluxes 
can be used as a proxy for AGN activity, as mid-IR emission correlates with X-ray 
brightness over a wide range of luminosities \citep{ags11,mlp12}. In particular the W4 filter at 24$\mu$m 
is not affected by dust absorption. As a comparison sample we choose the 887 type-2 quasars 
from \cite{rzs08}, 104 of which have similar redshifts as our GBs ($0.25\leq z \leq 0.35$). The 
de-redshifted mid-IR spectra of our GBs are very red, following a power-law with index 
$\langle a_\lambda\rangle=1.99\pm0.35$, showing that the emission is likely of nuclear origin as compared to 
star-formation. The 104 type-2 comparison quasars also have red spectra, albeit with
a lower slope of $\langle a_\lambda\rangle=1.59\pm0.43$. The null hypothesis that both samples have the 
same parent SED distribution is rejected on the 5\% level based on the Kolmogorov-Smirnov test. The 24$\mu$m 
luminosities of the GBs and the comparison sample are similar though, thus GBs might simply be particularly 
dusty objects.

\begin{figure}[t]
\includegraphics[width=1.0\hsize]{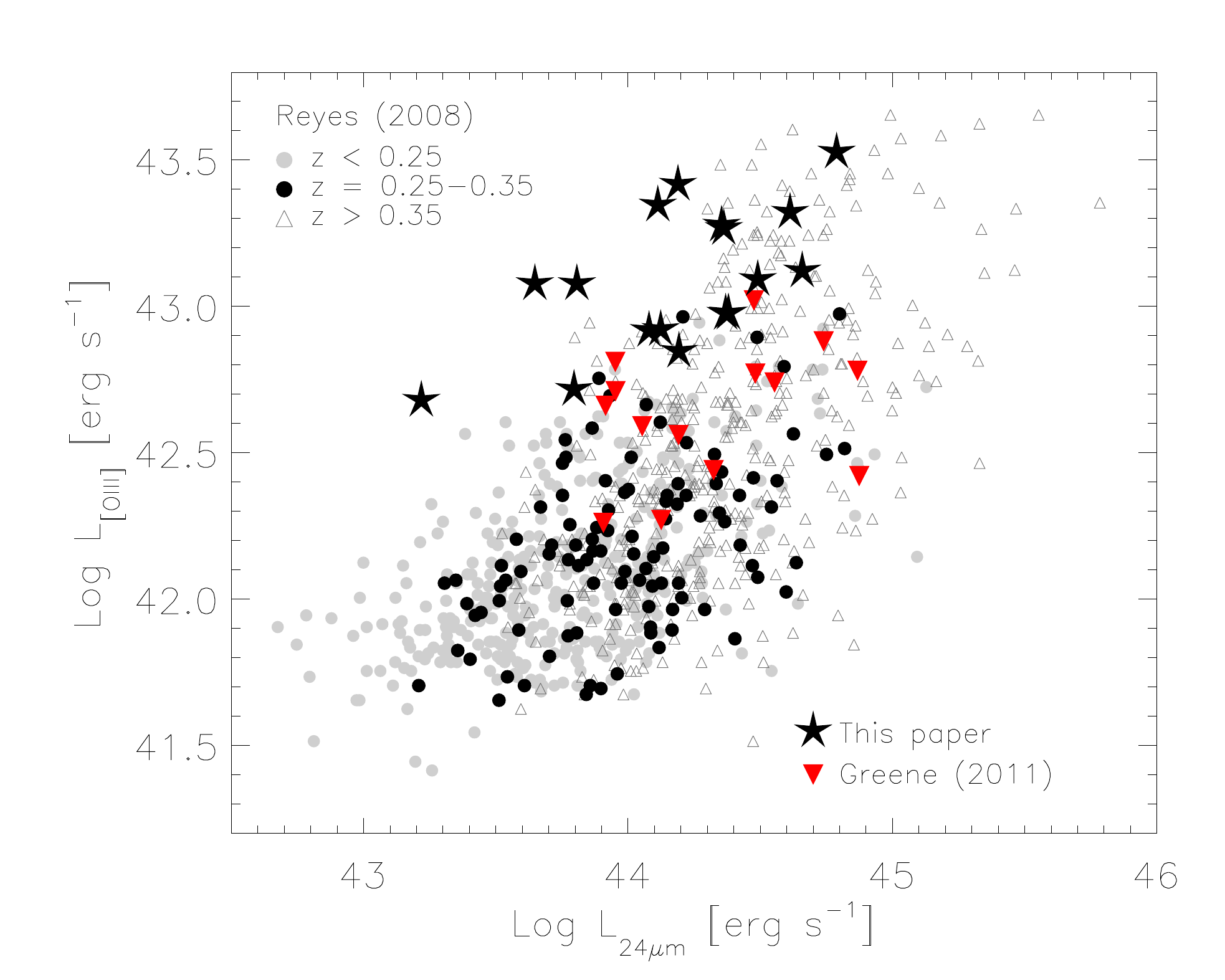}
\caption{\label{OIII-24mu}{[\ion{O}{III}]$\lambda5008$ vs.\ WISE 24$\mu$m luminosities. GBs are shown by asterisks,
and the red triangles represent the luminous obscured quasars studied by \cite{gzh11}. The other data points
show the type-2 AGN sample from \cite{rzs08} for three different redshift bins, with black dots
covering the same redshift range as GBs. At a given $L_{\rm 24\mu m}$, GBs have $5-50$ times
higher $L_{\rm [\ion{O}{III}]}$ than other type-2 AGN.}}
\end{figure}

In Fig.~\ref{OIII-24mu} we plot $L_{\rm [\ion{O}{III}]}$ versus $L_{\rm 24\mu m}$ for the obscured quasars of \cite{rzs08} 
and \cite{gzh11}. As our spectroscopic survey has not been flux calibrated, we have no direct measurements of 
$L_{\rm [\ion{O}{III}]}$ for the GBs. However we do know that the [\ion{O}{III}] equivalent widths are comparable to that 
of J2240, and that the spectra are generally similar. Assuming that [\ion{O}{III}]$\lambda5008$ contributes the same 
fraction to the total $r$-band flux as for J2240 (37\%), we overplot the GBs in Fig.~\ref{OIII-24mu}. The 
24$\mu$m luminosities for GBs and quasars in the same redshift range (black dots) are indistinguishable. However
[\ion{O}{III}] luminosities of the GBs are $5-50$ times higher than expected from their mid-IR emission. Since the latter 
mainly originates from the compact dusty torus, this indicates that the current AGN activity is 
too low to explain the NLR luminosity. The NLR may therefore reflect an earlier, more active state,
that subsided significantly in much less than a light crossing time. We return to this light echo hypothesis 
below.

\section{\label{discussion}Discussion}
\subsection{Main observational facts}
\subsubsection{AGN character and morphology}
The galaxy-wide and ultra-luminous NLR in J2240 is powered by an AGN, heating the medium and low
ionization zones to $15000-18000$ K and 13000 K, respectively. The Disk has higher temperature 
than the Cloud. The flux of the auroral lines is sufficiently high to facilitate more reliable 
temperature maps with deeper data. Studying the profile of the continuum emission within $\pm5$ 
kpc of the nucleus we find a disk-like Sersic index of 0.85 and a disk scale length of 3 kpc, hence 
ruling out an elliptical host galaxy.

We find large amounts of dust as expected for type-2 AGN. Dust is distributed within the NLR, and also 
in a large foreground patch (Fig.~\ref{extinctionall}). Reddening variations are high, and small 
systematics due to the unknown X-ray spectrum may still be present. BPT line 
diagnostics (Fig.~\ref{linediagnostics}) do not reveal significant star formation.

J2240 is dynamically complex, showing 7 distinct [\ion{O}{III}] peaks scattered over 30 kpc 
and 700 km s$^{-1}$ (Fig.~\ref{lines_all}). A rotational component is well detected in the Disk. A 
typical ENLR stretches over 26 kpc north-east and 42 kpc south-west from the nucleus (Fig.~\ref{halo}). 
It can be seen in all strong lines such as [\ion{O}{II}], [\ion{O}{III}], [\ion{Ne}{III}], [\ion{N}{II}], [\ion{S}{II}] and H$\alpha$, showing
that J2240 is embedded in a rotating and quiescent, yet highly excited, bubble of gas.

\subsubsection{Shock fronts and jets}
Electron density (lower right in Fig.~\ref{abu_ion_dens}) shows significant fluctuations. While the Cloud 
is mostly in the low-density limit of the [\ion{S}{II}] probe, $n_{\rm e}$ rises sharply to $650$ cm$^{-3}$ in the 
Center. This central peak has a smaller FWHM ($100$ km s$^{-1}$) than that of emission lines 
($200-500$ km s$^{-1}$), and it is offset by 1 kpc from the nucleus. Several blue- and redshifted 
secondary peaks exist within $\pm3$ kpc of the nucleus, the redshifted ones forming a chain where 
velocity increases with nuclear distance. Peaks also exist at larger separation in the Cloud. Shock fronts, 
interactions with a jet, or mergers may cause the peaks. Indeed [\ion{O}{I}]$\lambda6302$ is seen
throughout the NLR, verifying the global presence of shocks or ionization fronts. Since J2240 is radio-weak,
strong constraints for a jet are currently unavailable. The barely resolved radio contours in the VLA FIRST 
data are tentatively elongated along the major axis, thus a jet may explain at least some of the NLR 
shape.

The ionization parameter is strongly enhanced in the Center and dynamically broader than the 
density. The highest ionization values are offset with respect to the nucleus by a similar amount as that
observed for $n_{\rm e}$. A blue-shifted secondary peak is found near the nucleus with a relative 
velocity of \mbox{$-120$} km~s$^{-1}$, together with two more ionization peaks towards the Cloud, one of 
which in the area of highest extinction. Relative velocities are $+20$ and $+180$ km~s$^{-1}$. These secondary 
peaks may be the result of shocks, or point towards a second AGN which may either be deeply 
buried or merely missed by our long slit observation.

\subsubsection{Globally disturbed gas kinematics}
The [\ion{O}{II}] velocity profile is flat with ${\rm FWHM}=534\pm69$ km s$^{-1}$, and its peak coincides with the 
nucleus. Other lines, such as H$\alpha$, [\ion{O}{III}] and combined [\ion{S}{II}]$\lambda6718,33$, have almost identical 
profiles and larger variations in width than [\ion{O}{II}] ($200-550$ km s$^{-1}$; Fig.~\ref{lineprofile}). 
The largest FWHMs of these lines are found $1-3$ kpc north-east of the nucleus, where the medium ionization 
lines are bright, and $n_{\rm e}$ and $U$ the highest. The Cloud shows a secondary maximum in FWHM, and a 
smaller one exists between the Cloud and the nucleus. All velocity profiles drop to 
${\rm FWHM}\lesssim50$ km s$^{-1}$ beyond a $10-14$ kpc radius, typical for an ENLR. Within that radius, 
maximum velocities (taken at 20\% of the line profile) are nearly the same as the measured FWHM. Velocities 
exceed $1000$ km s$^{-1}$ near the nucleus, but these components are rather weak. Figure 
\ref{lineprofile} shows that neither the gas emission nor its velocity follow the stellar/continuum 
emission. Therefore the gas in J2240 is kinematically disturbed out to $10-14$ kpc radius.

\subsubsection{GB pilot survey}
Our ongoing spectroscopic pilot survey of GB candidates has confirmed at least 18 galaxies with NLRs 
$15-20$ kpc in size, and those covered by the VLA FIRST survey are all radio-quiet. 30\% of 
the GBs have close neighbors with yet unknown redshifts. In several cases [\ion{O}{III}] emission is 
dynamically perturbed. High-resolution spectroscopy in better seeing conditions than currently available 
may reveal systems with dynamic complexities as high as in J2240. All confirmed GBs are type-2 AGN, 
as broad-line components have not been identified. Upper limits to the fluxes of possible broad-line regions
were not obtained as the spectra were taken in non-photometric conditions and no flux standards were observed. 
[\ion{O}{III}] luminosities in GBs are on average one order of magnitude higher than expected from mid-IR emission, a 
sign of recently subsided AGN activity.

\subsection{Comparison with other NLRs}
\subsubsection{Space density and [OIII] luminosities}
In the 14500 deg$^2$ of SDSS-DR8 we find only 17 GBs of $0.12<z<0.36$. This redshift range corresponds to a 
co-moving volume of 11.1 Gpc$^3$. The space density of GBs is thus 4.4 Gpc$^{-3}$, consistent with
the upper end of the [\ion{O}{III}] luminosity function for type-2 quasars published by \cite{rzs08}. We have neglected 
the $0.39<z<0.69$ range in this consideration due to the contamination with star-forming galaxies.

With $26\times44$ kpc the NLR in J2240 is much larger than typical NLRs \citep[$0.1-5$ kpc,][]{bjk06}. 
Approximately 99\% of the [\ion{O}{III}] flux are contained within a radius of 25 kpc. J2240 falls within the 
scatter of the [\ion{O}{III}] 
size-luminosity relation observed for other Seyfert galaxies and quasars \citep{sda03,gzh11}. 
However, its total [\ion{O}{III}]$\lambda5008$ luminosity of $5.7\times10^{43}$ erg$\,$s$^{-1}$ is between 2 and 30 times 
higher than the maximum such value in the AGN samples of \cite{bmz10}, \cite{gea10} and \cite{lhp11}. It also 
exceeds the [\ion{O}{III}] luminosities of the high-z ULIRGs of \cite{has12} by factors $2-12$. It is one of the 
brightest objects amongst the 887 type-2 quasars of \cite{rzs08}. More luminous objects are only known at much 
higher redshifts, such as 2QZJ002830.4-281706 at $z = 2.4$ with $L_{\rm[\ion{O}{III}]}=2.6\times10^{44}$ erg s$^{-1}$ 
\citep{cmm12}. The [\ion{O}{III}] luminosities of the other GBs in our sample have so far only been estimated from their 
broad-band fluxes, but we showed they must also be in the range of several $10^{43}$ erg s$^{-1}$.

\subsubsection{Radio-loud quasars and ULIRGs}
EELRs are often found around radio-loud QSOs (Mrk 1014 is an exception, being radio-quiet). \cite{fus09} 
identify compact clouds moving with high velocities ($\sim500$ km s$^{-1}$), yet having low line widths 
($30-100$ km s$^{-1}$). They do not show morphological links with their host galaxies. EELRs are probably 
formed during mergers, which start the QSO engine that ionizes the gas and blasts it into the outer 
surroundings.

J2240 is different. Firstly, its host galaxy is radio-weak or -quiet. Secondly, the NLR is at 
least one order of magnitude brighter than the EELRs described in \cite{fym12}. Thirdly, J2240 shows high 
line widths ($200-500$ km s$^{-1}$) and low radial velocities ($\sim200$ km s$^{-1}$), opposite of what we see 
in EELRs. This may be an effect of incomplete sampling by the long slit, as we might have missed the high 
velocity components. Or the gas clouds move predominantly perpendicular to the line of sight such that we 
do not see the full velocity vector. Lastly, the NLR embeds the host galaxy, thus establishing a morphological
link. The same differences are also found when comparing GBs to the galaxy-scale emission line regions observed 
in powerful radio galaxies and some ULIRGs in the young Universe \citep[$z=2-3$;][]{nld08,has12}. This holds 
for the other GBs as well given our current data. 

\subsubsection{Current light echo samples}
J2240 does not appear to be a typical light echo, such as observed in Hanny's Voorwerp \citep{lsk09}, 
or those described by \cite{kcb12}. There are similarities though. Firstly, our BPT diagrams 
(Fig.~\ref{linediagnostics}) are fully compatible with the ones of \cite{kcb12}. Secondly, interaction 
with neighboring galaxies are frequently seen in the sample of \cite{kcb12}, and companions
are also observed for the GBs. 

Nevertheless, typical [\ion{O}{III}] luminosities for these light echos are $2$ orders of magnitude lower than for GBs.
This is also evident in the SDSS images, which show the stellar body of these galaxies well, whereas for GBs
the stellar emission is overwhelmed by the NLR. The survey of \cite{kcb12} 
gets insensitive at redshifts $z\gtrsim0.1$, as the investigated features become too faint. Our GB sample on the 
other hand, extracted from the same data base, does not reveal any GBs with $z<0.19$ (we are sensitive 
down to $z=0.12$). Shifting our SQL filter to even lower redshifts did not reveal further candidates. 
This is consistent for two reasons. First, a galaxy like J2240 at redshifts less than 0.1 would be conspicuous
and likely picked up by earlier surveys. Second, the comoving volume within $z<0.1$ is just about 0.3 
Gpc$^3$. If the space density of GBs calculated above does not evolve between $z\sim0.3$ and today, then we 
expect only 0.44 GBs at $z<0.1$ within the 14500 deg$^2$ covered by SDSS-DR8. For comparison, 
\cite{kcb12} find $\sim100$ galaxies with possible light echos at $z<0.1$.

\subsection{Binary AGN or SMBH merger?}
Can the extraordinary properties of J2240 be explained by mergers? We observe globally disturbed gas 
kinematics, and the highly ionized compact [\ion{O}{III}] sources can indicate multiplicity \citep{cgs12}.
A tidally distorted neighboring galaxy (Fig.~\ref{image}) makes a multiple merger scenario plausible,
in which another, possibly gas-rich, galaxy is currently consumed by J2240. As \cite{lss12} demonstrate 
for wide separation binary AGN, the SMBH accretion rate is increased in such double systems as the merger 
process funnels more material towards the centers. Log([\ion{O}{III}]) increases by $0.7\pm0.1$ when decreasing the 
separation from 100 to 5 kpc in these systems. Even closer pairs likely have correspondingly higher [\ion{O}{III}] 
luminosities.

Are we witnessing some violent process during the final stages of an AGN merger? SMBHs are common in the 
centers of massive galaxies, thus galaxy mergers must also result in the coalescence of SMBHs. Accordingly, 
SMBH or AGN pairs should be common. However, with decreasing separation they are increasingly hard to 
identify. For example, numerous binary AGN with separations of tens of kpc are known, yet with 3.6\% 
their fraction among optically selected AGN is already small \citep{lss11}. \cite{slg11} and \cite{cgs12} 
find kpc binary AGN in galaxies with double-peaked [\ion{O}{III}] emission, however that feature is more commonly 
caused by gas kinematics and cannot be used as a reliable indicator for AGN binarity \citep{slg11,fym12}. 
X-ray, infrared or radio observations are needed to confirm such systems.

Only few binary AGN with even smaller separations are known \citep{rtz06,fwe11}, including a system with 
sub-pc scale and an orbital period of $\sim100$ years \citep{bol09}. \cite{ebh12} report several 
sub-pc candidates, however emphasizing that long-term monitoring is required to link observed line 
variability with orbital motions. From the X-ray perspective statistics are equally weak. For example, 
\cite{tsu12} find that only $0-8$\% of massive mergers actually harbor binary AGN. 

While the observational data base is poor, simulations of the pre-coalescence state of SMBH mergers have 
been carried out \citep{hoq10,kbb12,vvm12}. During the orbital decay from kpc to pc scales no processes 
are found that explain an AGN flaring up for $10^4-10^5$ years by $3-4$ orders of magnitude. Neither can 
current accretion models explain a shut-down on similarly short time scales 
\citep{sev10}.

It is also unlikely that we are observing some effect or aftermath of the actual coalescence of two SMBHs.
\cite{thm10} calculate the electromagnetic footprint of SMBH mergers. They find an increase of bolometric 
luminosity of about 10\% per year over time scales of years or decades, together with an increase of X-ray 
hardness. This is, however, much less than the light crossing time of a galaxy ($10^4-10^5$ years) needed 
to explain the size and luminosity of our NLRs.
 
\subsection{Light echos - quasars shutting down?}
The most prominent characteristics of the NLRs in our GBs are their large angular extent and high [\ion{O}{III}] 
luminosities. Certainly powerful AGN must be responsible for this, but they are not evident from the SDSS 
imaging. Either they are deeply buried, or their activity has steeply declined over time scales much less 
than the light crossing time of the NLR, in which case we are observing strong light echos.

To this end we have to show that the current AGN activity is much lower than expected from the overall 
[\ion{O}{III}] luminosity, and that we are not just observing very obscured nuclei. The best way of doing this is
to determine the current X-ray luminosities of GBs, and compare them to their [\ion{O}{III}] luminosity. In case of a
light echo, the X-ray output will not match the [\ion{O}{III}] luminosity. Using the best-fit relation between 
$L_X$ and $L_{\rm [\ion{O}{III}]}$ for type-2 quasars and Seyferts, we expect $L_X\sim 1\times10^{44}$ erg s$^{-1}$ in 
the $2-10$ keV range \citep{jph12,lhp09}. Note that, since ROSAT is only sensitive to soft energies, even 
the lowest plausible X-ray absorption is sufficient to account for the non-detection of J2240 by ROSAT.
Therefore we cannot constrain a possible quasar shutdown with the X-ray data currently available.

To overcome the lack of X-ray data we use the mid-IR emission as a proxy for AGN activity, as it is 
unaffected by dust absorption and emanates from the immediate, pc-scale AGN environment. Mid-IR emission
is tightly correlated with the X-ray luminosity in AGN \citep{ags11,mlp12}. We compare WISE 24$\mu$m 
luminosities with [\ion{O}{III}]$\lambda5008$ luminosities for a large sample of type-2 quasars and GBs. Any 
change in AGN activity in GBs will need about a galaxy's light-crossing time before it is fully reflected 
in the NLRs' properties. We find the [\ion{O}{III}] luminosities of the GBs to be $5-50$ times higher than expected 
from the mid-IR emission, strengthening the light echo scenario. If J2240 and the other GBs are indeed light 
echos, they are spectacular examples of powerful QSOs currently shutting down.

Note, however, that the [\ion{O}{III}] luminosities of our GBs surpass the highest [\ion{O}{III}] luminosities of the reference 
type-2 sample at the same redshift (Fig.~\ref{OIII-24mu}). If the light echo interpretation is correct, then 
one expects quasars at similar redshifts with suitable mid-IR fluxes that have not shut down yet. Only quasars at 
$z\gtrsim0.4$ match such high fluxes. We think that this is a selection effect, with GBs dropping out of the 
main SDSS spectroscopic target selection algorithm. The fact that our simple size and color selection of GBs 
yields about 95\% spurious sources shows that the combination of broad-band photometry and angular size is 
quite unusual. The SDSS algorithm designed to select sources for spectroscopic follow-up is optimized for a high 
success rate, allowing for only 5\% unusual sources to be included \citep[see e.g.][and references therein]{rzs08}. 
It is therefore not surprising that only one of our GBs has a SDSS-DR8 spectrum.

\subsection{Outlook}
Considering the intrinsic luminosity that must be responsible for the observed optical NLR, the hard X-ray 
continuum of AGN must be directly detectable, provided that the column density is not too great 
($N_{\rm H} \lesssim 10^{23}$ cm$^{-2}$). Even if the central engine is deeply obscured or truly hidden by
Compton-thick absorption, the characteristic Fe K$\alpha$ fluorescence line should be detectable.  
In this way a binary AGN with a few kpc separation can be directly confirmed through X-ray imaging,
or the shut-down time-scale of the obscured quasars in GBs constrained. X-ray observations will also 
determine the slope of the ionizing spectrum, improving photo-ionization models of these NLRs.

Additional work going beyond our simple long-slit spectroscopic analysis has to be done. A realistic 
model of the NLR requires the use of photo-ionization codes and knowledge about the X-ray properties. 
The optical spectral analysis must also be extended to the full body of J2240. Such investigation
may reveal ionizing sources missed by the long-slit observation, and provide better extinction, 
temperature and density maps. In addition we can better constrain the dynamics, determine the gas mass, 
and investigate whether GBs have significantly higher gas masses than other obscured quasars showing massive
outflows or large NLRs. To this end, GMOS IFU observations in good seeing conditions have 
been conducted of J2240 in 2012B at the Gemini Observatory. We have also applied for follow-up observations of 
several more GBs.

\acknowledgments
MS thanks Hai Fu, Tohru Nagao, Evanthia Hatziminaoglou and Bob Fosbury for helpful discussions 
on the subjects of AGN, NLRs and star-bursts, Peter Storey for sharing his insight about 
atomic spectra, Gyula Jozsa for his input about radio properties, Yuri Beletsky and ESO for great 
support with the observations, and the anonymous referee for his recommendations.

\textit{Author contributions:} MS discovered J2240, obtained, reduced and analyzed all data and wrote 
the manuscript. RD, NL and CW helped with the interpretation and background information. KH sifted 
through the SDSS data base, fine-tuned the GB SQL filter, and provided substantial language 
corrections. NL provided the X-ray related aspects.

The authors wish to recognize and acknowledge the very significant cultural role and 
reverence that the summit of Mauna Kea has always had within the indigenous Hawaiian
community. We are most fortunate to have the opportunity to conduct observations from 
this mountain. MS acknowledges support by the German Ministry for Science and Education 
(BMBF) through DESY under the project 05AV5PDA/3 and the Deutsche Forschungsgemeinschaft 
(DFG) in the frame of the Schwerpunktprogramm SPP 1177 `Galaxy Evolution'.

This publication makes use of data products from the Wide-field Infrared Survey Explorer, 
which is a joint project of the University of California, Los Angeles, and the Jet Propulsion 
Laboratory/California Institute of Technology, funded by the National Aeronautics and Space 
Administration.

This research has made use of the VizieR catalog access tool, CDS, Strasbourg, France. 
Funding for the SDSS and SDSS-II has been provided by the Alfred P. Sloan Foundation, 
the Participating Institutions, the National Science Foundation, the U.S. Department of
Energy, the National Aeronautics and Space Administration, the Japanese Monbukagakusho, 
the Max Planck Society, and the Higher Education Funding Council for England. The SDSS 
Web Site is \anchor{http://www.sdss.org/}{http:www.sdss.org}. 

The SDSS is managed by the Astrophysical Research Consortium for the Participating 
Institutions. The Participating Institutions are the American Museum of Natural History,
Astrophysical Institute Potsdam, University of Basel, University of Cambridge, Case 
Western Reserve University, University of Chicago, Drexel University, Fermilab, the 
Institute for Advanced Study, the Japan Participation Group, Johns Hopkins University, 
the Joint Institute for Nuclear Astrophysics, the Kavli Institute for Particle
Astrophysics and Cosmology, the Korean Scientist Group, the Chinese Academy of Sciences 
(LAMOST), Los Alamos National Laboratory, the Max-Planck-Institute for Astronomy (MPIA), 
the Max-Planck-Institute for Astrophysics (MPA), New Mexico State University, Ohio State
University, University of Pittsburgh, University of Portsmouth, Princeton University, 
the United States Naval Observatory, and the University of Washington.

{\it Facilities:} \facility{VLT:Kueyen}, \facility{VLT:Antu}, \facility{CFHT}, 
\facility{Gemini:South}

\appendix
\onecolumn

\section{\label{apdx1} Further J2240-0927 data}
\begin{table}
\caption{\label{lineflux}
  Fluxes (corrected for galactic extinction) and luminosities of J2240 for selected emission lines, and 
  their relative strength with respect to H$\beta$. Total values are corrected for slit losses by seeing 
  and for the fact that the slit is smaller than the object. The correction factor used is $1.9\pm0.3$, 
  its uncertainty conservatively estimated.}
\begin{tabular}{lcccc}
  \noalign{\smallskip}
  \hline 
  \hline 
  \noalign{\smallskip}
  Line & $I^{\rm slit}$ & $L^{\rm slit}$ & $L^{\rm tot}$ & $I(\lambda)/I({\rm H}\beta)$ \\
       & [$\times10^{-16}\ergsm$] & [$\times10^{40} \ergs$] & [$\times10^{40} \ergs$] & \\
  \noalign{\smallskip}
  \hline 
  \noalign{\smallskip}
  H$\alpha$                     & $338.6\pm2.8 $ & $1200.2\pm9.9 $ & $2280\pm360$ & 4.25 \\
  H$\beta$                      & $ 79.7\pm1.2 $ & $ 282.4\pm4.1 $ & $ 536\pm85 $ & 1.00 \\
  H$\gamma$                     & $ 31.6\pm0.9 $ & $ 111.9\pm3.3 $ & $ 212\pm34 $ & 0.40 \\
  H$\delta$                     & $ 12.8\pm0.3 $ & $  45.6\pm1.1 $ & $  86\pm13 $ & 0.16 \\
  \ion{He}{I}$\,\lambda$5877            & $  7.6\pm0.1 $ & $  26.7\pm0.6 $ & $  50\pm8  $ & 0.09 \\
  \ion{He}{II}$\,\lambda$4687           & $ 10.6\pm0.4 $ & $  37.5\pm1.3 $ & $  70\pm11 $ & 0.13 \\
  \ion{He}{I}$\,\lambda$10833           & $ 23.7\pm0.8 $ & $  84.0\pm2.9 $ & $ 159\pm25 $ & 0.30 \\
  ${\rm [\ion{O}{I}]}\,\lambda$6302     & $ 44.0\pm0.6 $ & $ 156.0\pm2.1 $ & $ 296\pm46 $ & 0.55 \\
  ${\rm [\ion{O}{II}]}\,\lambda$3727    & $296.6\pm4.1 $ & $1051.3\pm14.7$ & $1997\pm316$ & 3.72 \\
  ${\rm [\ion{O}{III}]}\,\lambda$4364   & $ 10.3\pm0.5 $ & $  36.4\pm1.9 $ & $  69\pm11 $ & 0.13 \\
  ${\rm [\ion{O}{III}]}\,\lambda$5008   & $841.5\pm14.0$ & $2982.7\pm49.7$ & $5667\pm899$ & 10.56 \\
  ${\rm [\ion{N}{II}]}\,\lambda$6586    & $187.7\pm1.6 $ & $ 665.3\pm5.5 $ & $1264\pm199$ & 2.36 \\
  ${\rm [\ion{S}{II}]}\,\lambda$7618,33 & $168.1\pm1.6 $ & $ 595.6\pm5.4 $ & $1131\pm178$ & 2.11 \\
  ${\rm [\ion{S}{III}]}\,\lambda$9071   & $ 29.6\pm0.9 $ & $ 104.9\pm3.1 $ & $ 199\pm32 $ & 0.37 \\
  ${\rm [\ion{S}{III}]}\,\lambda$9533   & $107.2\pm2.6 $ & $ 379.8\pm9.2 $ & $ 721\pm115$ & 1.34 \\
  ${\rm [\ion{Ne}{III}]}\,\lambda$3869  & $ 56.3\pm0.9 $ & $ 199.6\pm3.0 $ & $ 379\pm60 $ & 0.71 \\
  ${\rm [\ion{Ne}{V}]}\,\lambda$3427    & $ 17.4\pm0.4 $ & $  61.8\pm1.6 $ & $ 117\pm18 $ & 0.22 \\
  ${\rm [\ion{Mg}{II}]}\,\lambda$2799   & $ 19.5\pm0.5 $ & $  69.1\pm1.9 $ & $ 131\pm21 $ & 0.24 \\
  ${\rm [\ion{Ar}{III}]}\,\lambda$7138  & $  8.9\pm0.1 $ & $  31.4\pm0.6 $ & $  59\pm9  $ & 0.11 \\
  \hline 
  \noalign{\smallskip}
\end{tabular}
\end{table}

\begin{figure}
\includegraphics[width=1.0\hsize]{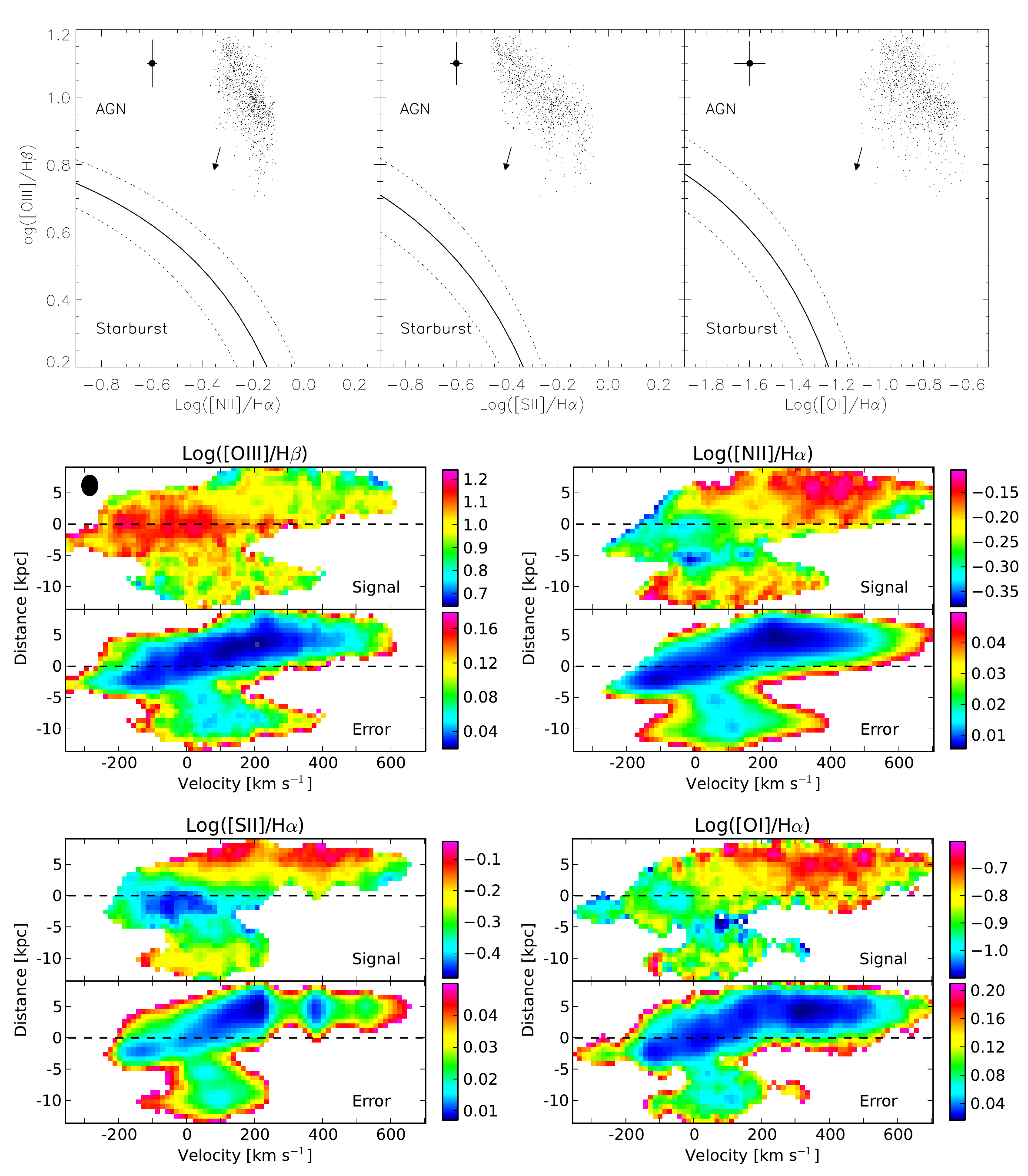}
\caption{\label{linediagnostics}BPT diagrams. Top: The solid and dotted lines represent the 
  model division between the two galaxy types and its uncertainty \citep{kds01}. Data points show
  individual pixels in the NLR, their mean errors are indicated. The NLR is evidently powered by 
  an AGN. The maximum possible effect of stellar absorption is indicated by the arrows. Middle and 
  bottom: Spatially and spectrally resolved BPT diagrams, based on the same data points as shown 
  in the top row.}
\end{figure}

\begin{figure}
\includegraphics[width=1.0\hsize]{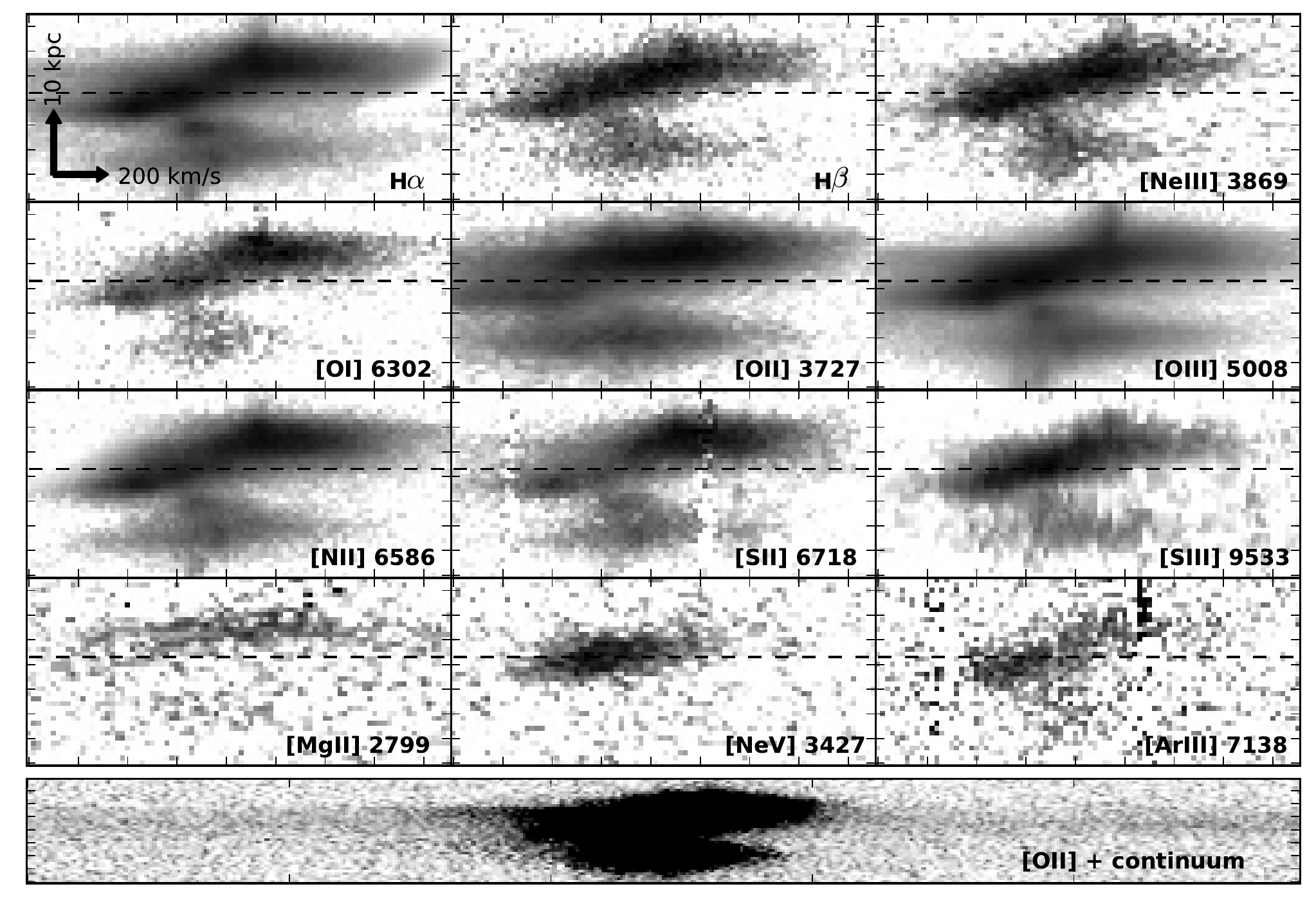}
\caption{\label{lines_all}Top 4 rows: Selected emission lines (continuum 
  subtracted, and projected to a common reference frame) in J2240. A logarithmic scaling (different 
  between lines) has been chosen for better visualization. The dashed line indicates the spatial 
  position of the nucleus (position of highest continuum level), i.e.\ the spatial direction is 
  oriented along the vertical axis. Wavelength increases from left to right. Spatial and velocity scales
  are indicated in the upper left panel. The area displayed for each line is slightly larger than that 
  used for the 2D spectral analysis plots, in order to show more details.
  Bottom: For comparison, we show a linearly scaled larger part of the spectrum before 
  continuum subtraction, centered on [\ion{O}{II}]$\lambda3727$. This gives an idea about the strength 
  and extent of the emission lines (see also Fig.~\ref{emissionlines}).}
\end{figure}

\begin{figure}
\includegraphics[width=1.2\hsize,angle=90]{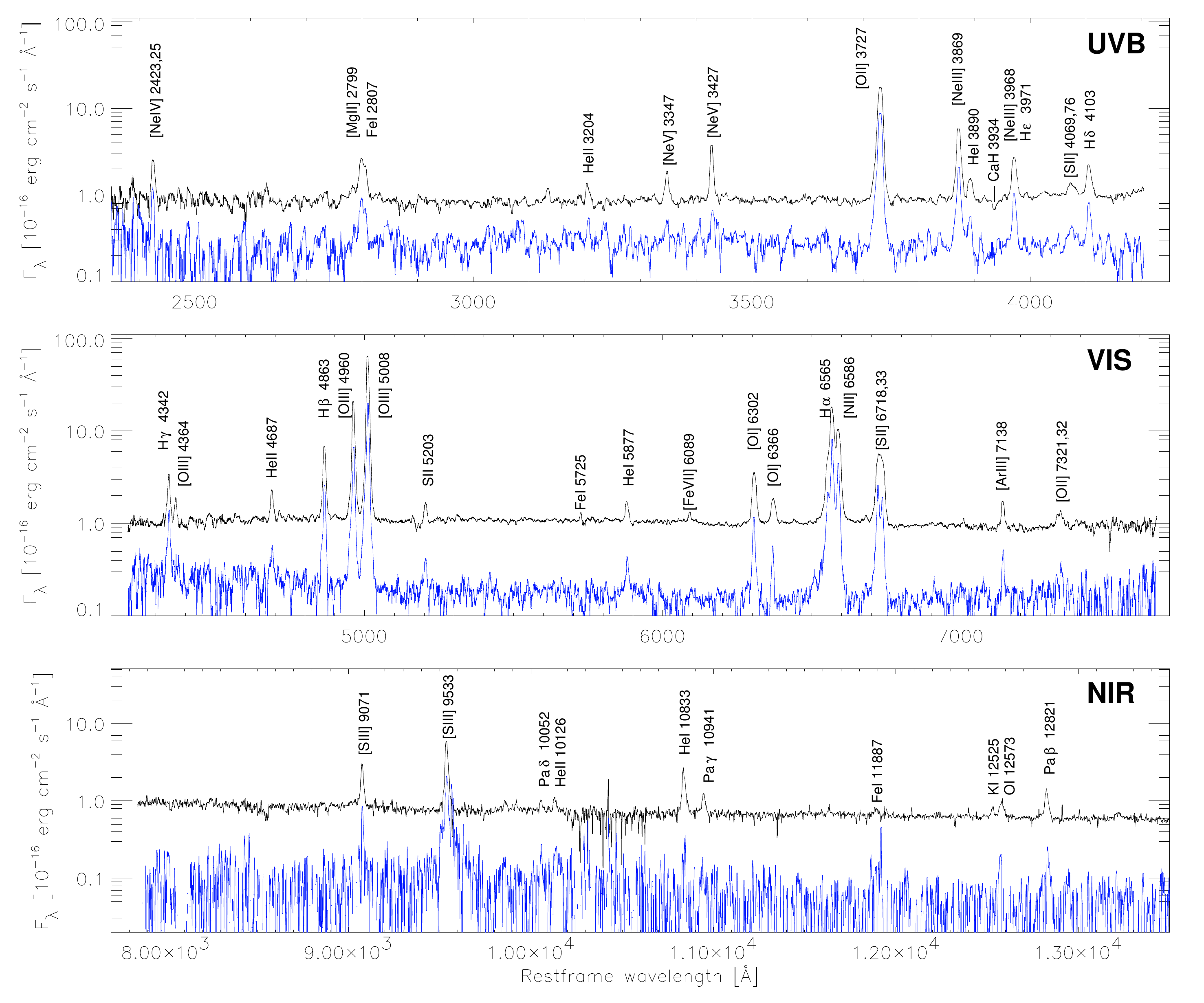}
\caption{\label{emissionlines}Rest-frame 1D XSHOOTER spectra of J2240, uncorrected for galactic extinction. 
  The black line represents the galaxy center, integrated within $\pm4.5$ kpc of the nucleus. The blue line has 
  been integrated over 7.6 kpc centered on the Cloud. Note the great similarity between the two spectra. For 
  visualization purposes data have been filtered with a 7\AAb wide median kernel, thus the actual resolution 
  is 48 (12) times higher than shown for the UVB/VIS (NIR) channels, respectively.}
\end{figure}
\clearpage

\section{\label{gbsample}The GB sample}
The following SQL filter was used to retrieve the initial selection of 376 GB candidates from 
SDSS-DR8\footnote{\anchor{http://skyserver.sdss3.org/dr8/en/tools/search/sql.asp}
{http://skyserver.sdss3.org/dr8/en/tools/search/sql.asp}}, including poststamps for a visual 
cross-check. Only $\sim5$\% of the objects are 
retained as genuine GB candidates after visual inspection. We experimented with various
parameters, such as the {\tt Clean} flag, but found that these were prone to exclude genuine
objects (amongst others J2240).

\vspace{0.3cm}
{\tt 
\noindent SELECT\\
\hspace*{1cm}  ra, dec, objID, u,g,r,i,z, petrorad\_r,\\
\hspace*{1cm}  '<a href=http://cas.sdss.org/dr3/en/tools/chart/navi.asp?ra='+\\
\hspace*{1cm}  cast(ra as varchar(10))+'\&dec='+cast( dec as varchar(10)) + '>'+\\
\hspace*{1cm}  '<img src="http://skyservice.pha.jhu.edu/dr8/ImgCutout/getjpeg.aspx?ra='+\\
\hspace*{1cm}  cast(ra as varchar(15))+'\&dec='+cast(dec as varchar(15))+\\
\hspace*{1cm}  '\&scale=0.40\&width=120\&height=120\&opt="/> ' as pic\\
FROM\\
\hspace*{1cm}  Galaxy\\
WHERE\\
\hspace*{1cm}  ((r <= 20.5)\\
\hspace*{1cm}  and (r >= 17)\\
\hspace*{1cm}  and (u - r <= 5)\\
\hspace*{1cm}  and (r - i <= -0.2)\\
\hspace*{1cm}  and (r - z <= 0.6)\\
\hspace*{1cm}  and (g - r >= r - i + 0.5)\\
\hspace*{1cm}  and (u - r >= 2.5 * (r - z))\\
\hspace*{1cm}  and (g - r > 1.0)\\
\hspace*{1cm}  and (petrorad\_r > 2)\\
\hspace*{1cm}  and (psfmagerr\_g < 0.04)\\
\hspace*{1cm}  and (psfmagerr\_r < 0.04)\\
\hspace*{1cm}  and (psfmagerr\_i < 0.04))
}

\begin{sidewaystable}[t]
\caption{\label{gbsample-table}GB sample. $R_{r,i}$ are the SDSS Petrosian radii. All but two redshifts originate 
from our GMOS pilot survey. High values of ${\rm log}({\rm [\ion{O}{III}]}/{\rm H\beta})\sim1$ confirm the presence of an 
AGN. The last two data columns contain VLA FIRST radio and WISE 24$\mu$m fluxes. No number indicates that the
source is either outside the FIRST survey area, or no spectra have been taken. Object \#016 is the main target 
analyzed in this paper. Spectroscopic data for \#022 were taken from archival GMOS-North observations, executed by
N. Zakamska (Gemini program GN-2010B-C-10). }

\begin{tabular}{lccccccccl}
  \noalign{\smallskip}
  \hline 
  \hline 
  \noalign{\smallskip}
  \# & Name & SDSS objID & $r$ [mag] & $R_{r}\,[^{\prime\prime}]$ & $z$ & ${\rm log}\left(\frac{{\rm [\ion{O}{III}]}}{{\rm H}\beta}\right)$ & 
  $F_R$ [mJy] & $F_{24\mu{\rm m}}$ [mJy] & Comments\\
  \noalign{\smallskip}
  \hline 
  \noalign{\smallskip}
001 & SDSS J002016.4$-$053126 & 1237679077517557845 & 18.3 & 2.1 & 0.334   & $1.305\pm0.005$ & 9.38       & 11.7 & \\
002 & SDSS J002434.9$+$325842 & 1237676441460474246 & 18.2 & 2.6 & 0.293   & $1.246\pm0.003$ & \nodata    & 25.4 & \\
003 & SDSS J011133.3$+$225359 & 1237666091128914338 & 19.1 & 2.1 & 0.318   & $1.243\pm0.016$ & \nodata    & 23.1 & \\
004 & SDSS J011341.1$+$010608 & 1237666340800364769 & 18.5 & 2.6 & 0.281   & $1.043\pm0.006$ & 1.18       & 39.7 & SDSS redshift\\
005 & SDSS J015930.8$+$270302 & 1237680284389015833 & 18.9 & 2.6 & 0.278   & $1.194\pm0.004$ & \nodata    & 18.1 & \\
006 & SDSS J115544.5$-$014739 & 1237650371555229774 & 17.9 & 2.4 & 0.306   & $1.163\pm0.006$ & undetected & 16.9 & \\
007 & SDSS J134709.1$+$545311 & 1237661386529374363 & 18.7 & 2.0 & \nodata & \nodata & 3.67               &  5.3 & \\
008 & SDSS J135155.4$+$081608 & 1237662236402647262 & 19.0 & 2.1 & 0.306   & $1.161\pm0.002$ & undetected & 25.7 & \\
009 & SDSS J144110.9$+$251700 & 1237665442062663827 & 18.5 & 2.2 & 0.192   & $1.086\pm0.001$ & 2.28       & 19.6 & \\
010 & SDSS J145533.6$+$044643 & 1237655742407835791 & 18.5 & 2.0 & 0.334   & $1.151\pm0.003$ & undetected & 20.4 & \\
011 & SDSS J150420.6$+$343958 & 1237662306730639531 & 18.7 & 2.2 & \nodata & \nodata & undetected         &  7.6 & \\
012 & SDSS J150517.6$+$194444 & 1237667968032637115 & 17.9 & 2.4 & 0.341   & $1.131\pm0.001$ & 4.89       & 49.9 & \\
013 & SDSS J205058.0$+$055012 & 1237669699436675933 & 18.6 & 2.6 & 0.301   & $1.163\pm0.003$ & \nodata    & 49.5 & \\
014 & SDSS J213542.8$-$031408 & 1237680191506678389 & 19.2 & 2.0 & 0.246   & $1.108\pm0.004$ & \nodata    &  3.2 & \\
015 & SDSS J220216.7$+$230903 & 1237680306395415794 & 18.9 & 2.5 & 0.258   & $1.154\pm0.006$ & \nodata    & 24.8 & \\
016 & SDSS J224024.1$-$092748 & 1237656538051248311 & 18.3 & 3.4 & 0.326   & $0.960\pm0.003$ & 2.85       & 37.4 & Prototype; J2240\\
017 & SDSS J230829.4$+$330310 & 1237680503434445439 & 19.1 & 2.0 & 0.284   & $1.258\pm0.010$ & \nodata    & 13.9 & \\
  \hline
  \noalign{\smallskip}
  ID & RA (J2000) & DEC (J2000) & $i$ [mag] & $R_{i}\,[^{\prime\prime}]$ & $z$ & ${\rm log}\left(\frac{{\rm [\ion{O}{III}]}}{{\rm H}\beta}\right)$ & $F_R$ [mJy] & $F_{24\mu{\rm m}}$ [mJy] & Comments\\
  \noalign{\smallskip}
  \hline
  \noalign{\smallskip}
018 & SDSS J015629.0$+$174940 & 1237679459756015771 & 19.2 & 2.1 & \nodata & \nodata & \nodata            &   5.7 & \\ 
019 & SDSS J080001.8$-$095841 & 1237676676082173676 & 18.9 & 2.5 & 0.402   & $1.232\pm0.010$ & \nodata    &  14.7 & \\
020 & SDSS J090015.7$+$604704 & 1237663530800382081 & 16.9 & 2.6 & 0.082   & \nodata & 178.71   &  25.2 & \ion{H}{II} region\\
021 & SDSS J091353.4$+$601748 & 1237651274038837416 & 18.0 & 3.1 & 0.477   & $1.141\pm0.0027$ & 1.36      &  31.7 & \\
022 & SDSS J110140.5$+$400423 & 1237661966349893812 & 18.2 & 2.6 & 0.456   & $1.296\pm0.014$ & 17.38      &  18.8 & N. Zakamska\\
023 & SDSS J113818.9$+$060620 & 1237654606942699779 & 18.8 & 2.2 & 0.499   & $1.132\pm0.009$ & undetected &  undetected & \\
024 & SDSS J122140.0$+$190442 & 1237667914346791043 & 18.6 & 2.9 & 0.542   & $1.163\pm0.010$ & undetected & 30.3 & \\
025 & SDSS J161836.4$-$040942 & 1237668567712924379 & 17.6 & 2.2 & 0.211   & 
    $0.155\pm0.008$ & undetected & 268.4 & \ion{H}{II} region\\
026 & SDSS J172145.7$+$632233 & 1237671938731213392 & 18.9 & 2.2 & \nodata & \nodata & undetected & 5.9 & \\
027 & SDSS J191354.1$+$621147 & 1237671938736848903 & 18.8 & 2.2 & \nodata & \nodata & undetected & 32.7 & \\
028 & SDSS J230901.0$+$035742 & 1237679004484370740 & 18.5 & 2.3 & 0.485   & $1.188\pm0.007$ & undetected & 12.8 & \\
029 & SDSS J234130.3$+$031727 & 1237678598088425669 & 17.3 & 2.4 & 0.145   & $0.612\pm0.014$ & 851.95 & 121.8 & 
    \ion{H}{II} region; LINER\\
    & & & & & & & & & strong [\ion{O}{I}], weak [\ion{O}{III}]\\
  \hline
\end{tabular}
\end{sidewaystable}

\clearpage

\begin{figure}
  \includegraphics[width=1.0\hsize]{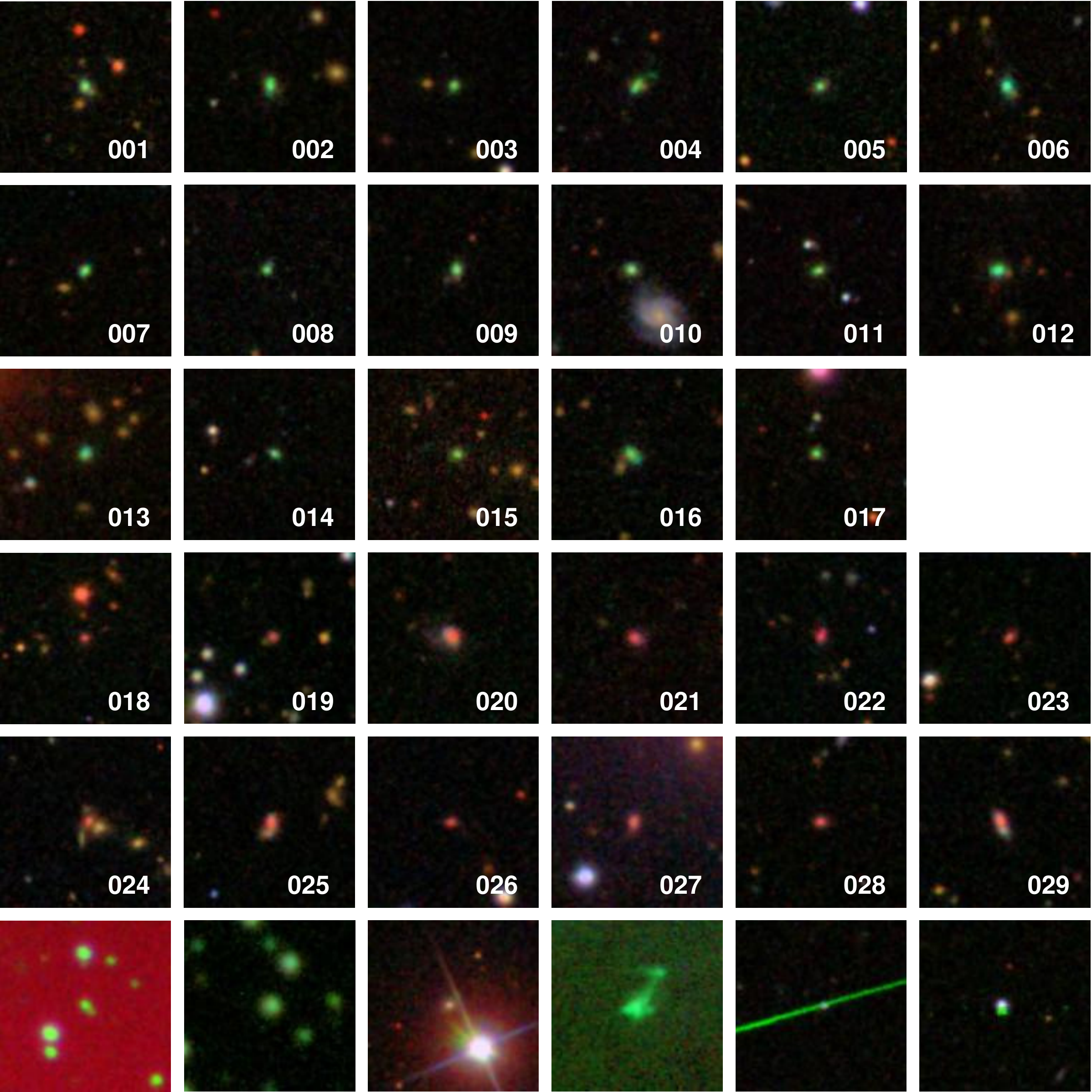}
  \caption{\label{gbsample-figs}{Top three rows: GBs with $0.12<z<0.36$
    (see also Table~\ref{gbsample-table}}). \#016 is J2240. Rows 4 and 5: Possible GBs
    with $0.39<z<0.69$. Bottom: common false positives, from left to right: bad background, 
    systematic photometry error or bad seeing, bright star/diffraction spike, Herbig-Haro 
    object, satellite, star near detector edge.}
\end{figure}

\begin{figure}[t]
  \includegraphics[width=1.0\hsize]{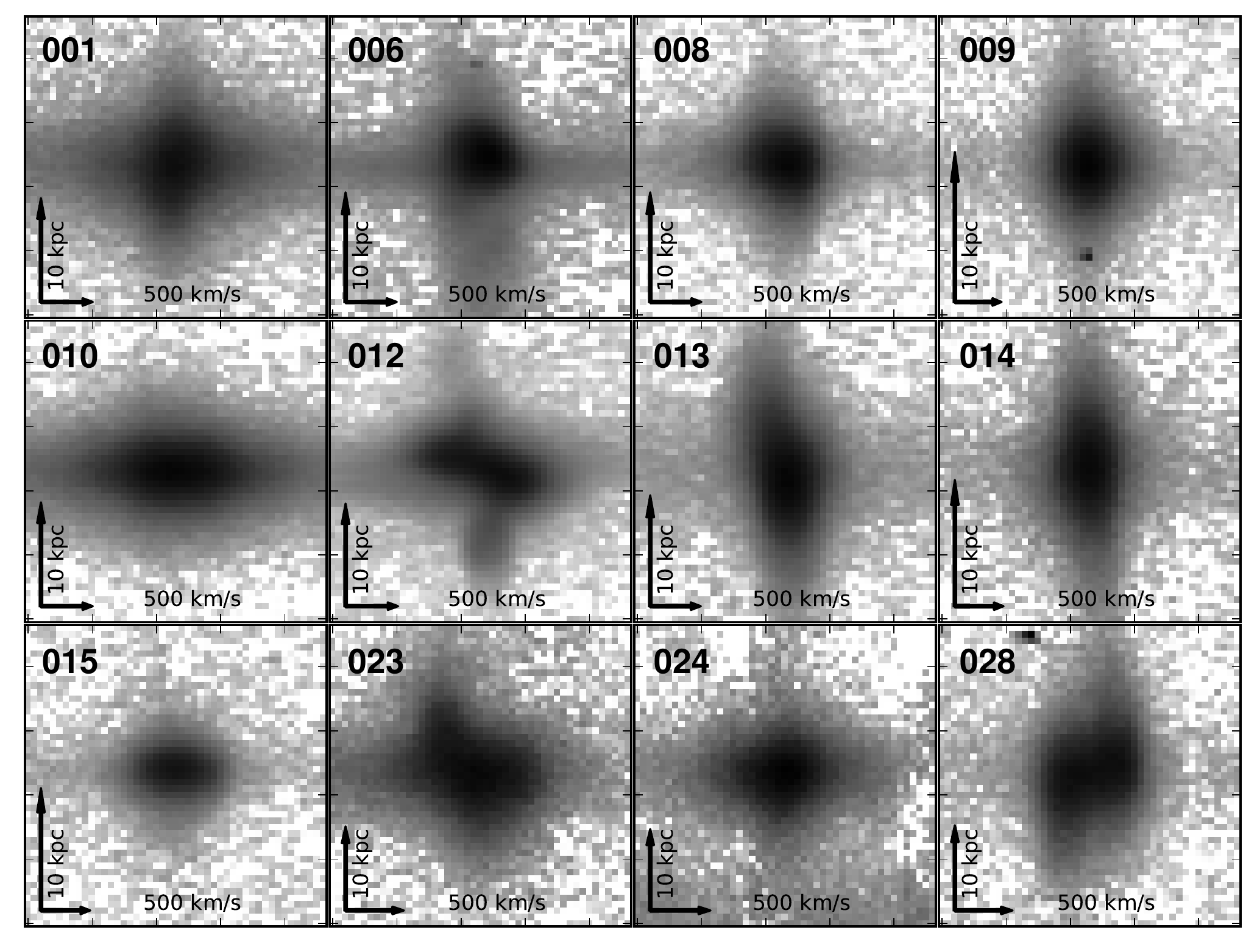}
  \caption{\label{gbsample_OIII}{[\ion{O}{III}]$\lambda5008$ (log-scaled) for some GBs. 
      The continuum has not been subtracted.}}
\end{figure}

\bibliography{mybib}

\end{document}